\documentclass[review]{elsarticle}
\usepackage[
    type={CC},
    modifier={by-nc-nd},
    version={4.0},
]{doclicense}

\usepackage{lineno,hyperref}

\usepackage{enumitem}
\usepackage{url}
\usepackage{glossaries}
\usepackage{graphicx}
\usepackage{subfig}
\usepackage{color}
\usepackage{todonotes}
\usepackage{ulem}
\usepackage{amssymb}

\journal{Astroparticle Physics}

\begin{document}

\newacronym{gf}{GF}{Geomagnetic Field}
\newacronym{eas}{EAS}{Extensive Air Showers}
\newacronym{cta}{CTA}{Cherenkov Telescope Array}
\newacronym{nsb}{NSB}{night-sky background}
\newacronym{mars}{MARS}{MAGIC Reconstruction Software}

% For the "sloppy" mode: http://tex.stackexchange.com/questions/89354/forcing-math-mode-to-be-on-the-same-line
\binoppenalty=10000
\relpenalty=10000

\newcommand{\tmtexttt}[1]{{\ttfamily{#1}}}
\newcommand{\apj}{Astrophysical Journal}
\newcommand{\aap}{A\&A}
\newcommand{\pasp}{Publications of the ASP}

\definecolor{navyblue}{rgb}{0.0, 0.0, 0.5}
\newcommand{\thadd}[1]{\textcolor{navyblue}{[TH: #1]}}

\begin{frontmatter}

\title{Monte Carlo Performance Studies for the Site Selection of the Cherenkov Telescope Array}

\author[inst-ifae]{T. Hassan\corref{cor1}}
       \ead{thassan@ifae.es}

\author[inst-lupm]{L. Arrabito}
\author[inst-mpik]{K. Bernl{\"o}hr}
\author[inst-lupm]{J. Bregeon}
\author[inst-ifae]{J. Cortina}
\author[inst-ifae]{P. Cumani}
\author[inst-inaf]{F. Di Pierro}
\author[inst-saoPaulo]{D. Falceta-Goncalves}
\author[inst-saoPaulo3]{R. G. Lang}
\author[inst-mpik]{J. Hinton}
\author[inst-slac]{T. Jogler}
\author[inst-desyz]{G.~Maier}
\author[inst-ifae]{A. Moralejo}
\author[inst-inaf2]{A. Morselli}
\author[inst-saoPaulo2]{C. J. Todero Peixoto}
\author[inst-slac]{M. Wood}

\address[inst-ifae]{Institut de Fisica d'Altes Energies (IFAE), The Barcelona Institute of Science and Technology, Campus UAB, 08193 Bellaterra (Barcelona) Spain}
\address[inst-desyz]{DESY, Platanenallee 6, D-15738 Zeuthen, Germany}
\address[inst-lupm]{Laboratoire Univers et Particules de Montpellier - UMR5299, Universit\'e de Montpellier - CNRS/IN2P3, Place Eug\`ene Bataillon - CC 72, 34095 Montpellier C\'edex 05 France}
\address[inst-mpik]{Max-Planck-Institut f{\"u}r Kernphysik, P.O. Box 103980, 
        D-69029 Heidelberg, Germany}
\address[inst-inaf]{INFN Sezione di Torino, Via P. Giuria 1, 10125 Torino, Italy}
\address[inst-inaf2]{INFN Roma Tor Vergata, Via della Ricerca Scientifica 1, 00133 Roma, Italy}
\address[inst-slac]{SLAC, Stanford, CA~94025, USA}
\address[inst-saoPaulo]{Universidade de S\~{a}o Paulo - Escola de Artes, Ci\^{e}ncias e Humanidades
Rua Arlindo B\'{e}ttio, no. 1000 - Ermelino Matarazzo - S\~{a}o Paulo - SP CEP: 03828-000}
\address[inst-saoPaulo2]{Universidade de S\~{a}o Paulo - Escola de Engenharia de Lorena, Departamento de Ci\^{e}ncias B\'{a}sicas e Ambientais - Estrada do Municipal do Campinho s/no. - Lorena - S\~{a}o Paulo - CEP: 12602-810}
\address[inst-saoPaulo3]{Instituto de F\'{i}sica de So Carlos, Universidade de S\~{a}o Paulo, Av. Trabalhador S\~{a}o-carlense, 400 - Pq. Arnold Schimidt, CEP: 13566-590, S\~{a}o Carlos, SP, Brazil}

\cortext[cor1]{Corresponding author}

\begin{abstract}

The \gls{cta} represents the next generation of ground-based instruments for very-high-energy (VHE) gamma-ray astronomy, aimed at improving on the sensitivity of current-generation experiments by an order of magnitude and providing coverage over four decades of energy. The current CTA design consists of two arrays of tens of imaging atmospheric Cherenkov telescopes, comprising Small, Medium and Large-Sized Telescopes, with one array located in each of the Northern and Southern Hemispheres. To study the effect of the site choice on the overall \gls{cta} performance and support the site evaluation process, detailed Monte Carlo simulations have been performed. These results show the impact of different site-related attributes such as altitude, night-sky background and local geomagnetic field on \gls{cta} performance for the observation of VHE gamma rays.

\end{abstract}

\begin{keyword}

Monte Carlo simulations \sep
Cherenkov telescopes \sep
IACT technique \sep
gamma rays \sep
cosmic rays

\end{keyword}

\end{frontmatter}

%%%%%%%%%%%%%%%%%%%%%%%%%%%%%%%%%%%%%%%%%%%%%%%%%%%%%%%%%%
\glsreset{cta}
\section{Introduction}
%\todo{ABSTRACT: atmospheric profile?}
As a result of the success of current imaging atmospheric Cherenkov telescopes (IACTs) and the improvement of the different technologies involved, the next generation of ground-based very-high-energy (VHE) gamma-ray detectors is under development. The \gls{cta}\footnote{\url{http://www.cta-observatory.org/}} \cite{CTA_concept, CTAICRC2015} will give deep and unprecedented insight into the non-thermal high-energy Universe scrutinising the gamma-ray sky from 20 GeV to 300 TeV, improving the sensitivity of current instruments by more than an order of magnitude.

In order to achieve these goals, the CTA Observatory will consist of two different sites, one in each Hemisphere, and telescopes of three different sizes: Large-Sized Telescopes (LSTs) \cite{LSTgamma} sensitive to the faint low-energy showers (below 200 GeV),  Medium-Sized Telescopes (MSTs) \cite{MSTgamma, SCTgamma} increasing the effective area\footnote{The effective area of the instrument is defined as the differential gamma-ray detection rate, $\frac{dN_{\gamma,det}}{dE}$, after all analysis cuts (see Sec. \ref{sec:analysis}), divided by the differential flux of incident gamma rays.} and the number of telescopes simultaneously observing each event within the CTA core energy range (between 100 GeV and 10 TeV) and Small-Sized Telescopes (SSTs) \cite{Montaruli:2015} spread out over several km$^2$ to increase the number of detected events at the upper end of the electromagnetic spectrum accessible to CTA (up to $\sim$ 300 TeV).

The proposed designs for the Northern and Southern observatories will make the full sky accessible with an improved sensitivity alongside better angular and energy reconstruction. The CTA Southern site, with an ideal location to observe the Galactic Center and a big fraction of the Galactic Plane, will be larger in order to measure the extremely low fluxes expected from these sources above 10 TeV. Its baseline design foresees 4 LSTs, 25 MSTs, and 70 SSTs \cite{Hassan-2015}. The Northern site, with a broader coverage of the extragalactic sky, will be more focused on the study of extragalactic objects and transient phenomena. The CTA Northern Hemisphere site is planned to be composed of 4 LSTs and 15 MSTs.

One of the advantages of such an extended telescope layout is that most of the detected events will be fully contained inside the area covered by the array. These so-called \textit{contained events} will be better sampled, providing an improved background rejection, better angular and energy resolution, and reduced energy threshold.

The criteria considered for the scientific site ranking by the CTA Consortium are costs, risks and scientific performance. Costs (including host premiums, available infrastructure, building and operation costs, taxes and fees) and risks (including economic and socio-political risks or environmental hazards) are not considered in the present paper. The scientific performance of a candidate site depends mainly on the average annual observing time (AAOT) and the performance per unit time\footnote{Throughout this work, the differential sensitivity in 50 hours of observation (defined in section \ref{siteParameters}) will be used as the main parameter describing an array performance per unit time.} (PPUT) of the array. The AAOT, mainly dependent on the site's weather conditions, was evaluated for each candidate site using various satellites and weather simulations \cite{weather_sims}, together with in-situ weather records. The AAOT is also beyond the scope of the paper.

All sites proposed to host such an ambitious project satisfy a list of geographical and atmospheric criteria. Sites were required to be located at medium latitudes, contain enough available area for the deployment of the telescopes layout, have a clean atmosphere with no obstacles blocking significant parts of the sky and tolerable annual ranges of temperature, wind and humidity.

This study focuses on the determination of the scientific performance per unit time of each proposed site, and evaluates through detailed MC simulations the effect of several site attributes like altitude, geomagnetic field and \gls{nsb} on the telescope layout performance. These site-related parameters have been widely studied by the current generation of IACTs, and are briefly described in the following section.

\section{Site parameters and CTA performance}
\label{siteParameters}

To optimise the CTA design, detailed Monte Carlo simulations have been performed to estimate its scientific performance (\cite{APP_CTA_MC, MC_ICRC:2013, MC_ICRC_site:2015,Hassan-2015}). Throughout this work, the differential sensitivity to point-like sources is the parameter used to evaluate CTA performance per unit time. The differential sensitivity, i.e.~minimum detectable flux from a steady, point-like gamma-ray source, calculated for a narrow energy range, depends on the collection area, angular resolution, and rate of background events surviving all gamma selection cuts.

IACTs capture images of the very short flashes (a few ns) of optical Cherenkov radiation caused by the charged particles generated within the extensive air showers (EASs) produced by VHE gamma and cosmic rays. Most of this light is emitted at an altitude of 5--10 km and propagates as a cone with a small opening angle (0.5--1 deg). At ground level, the shower results in a pool of light of $\sim$ 120 m radius centered at the \textit{core position}. As shown in Fig. \ref{fig:lateral}, the lateral distribution of the Cherenkov light emitted within the EASs (i.e. average Cherenkov photon density reaching ground as a function of the distance to the core) changes significantly with the energy of the primary particle. Captured images picture the emitted Cherenkov photons through the atmosphere projected within the line of sight of each IACT as elongated elliptical-shaped images. Then the primary particle is identified (as a gamma ray or background) and original direction reconstructed (with up to sub-arc-minute accuracy) using the orientation and shape of all recorded images of the EAS.

The considered CTA candidate sites are listed in Table \ref{table:sites}, together with some relevant site-related parameters. These parameters directly affect the performance of IACTs as they influence the development of the EASs \cite{impactAtmospherics}, modifying the Cherenkov light density at ground level. The main environmental parameters affecting the sensitivity of IACTs are the site altitude, the local geomagnetic field intensity and the \gls{nsb} level.

\subsection{Altitude}
\label{subsec:altitude}

\begin{figure}
\begin{center}
\centering\includegraphics[width=0.9\linewidth]{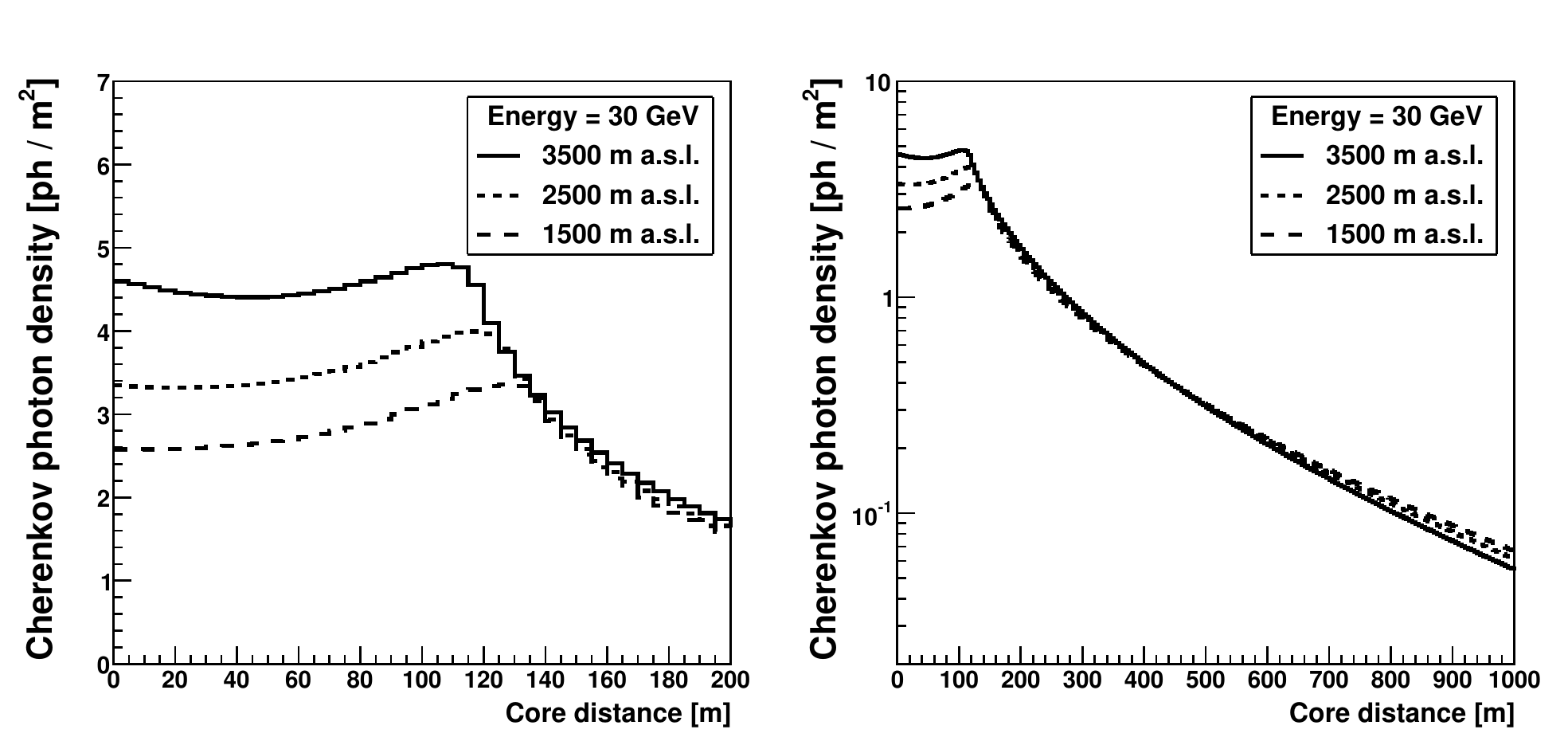}
\centering\includegraphics[width=0.9\linewidth]{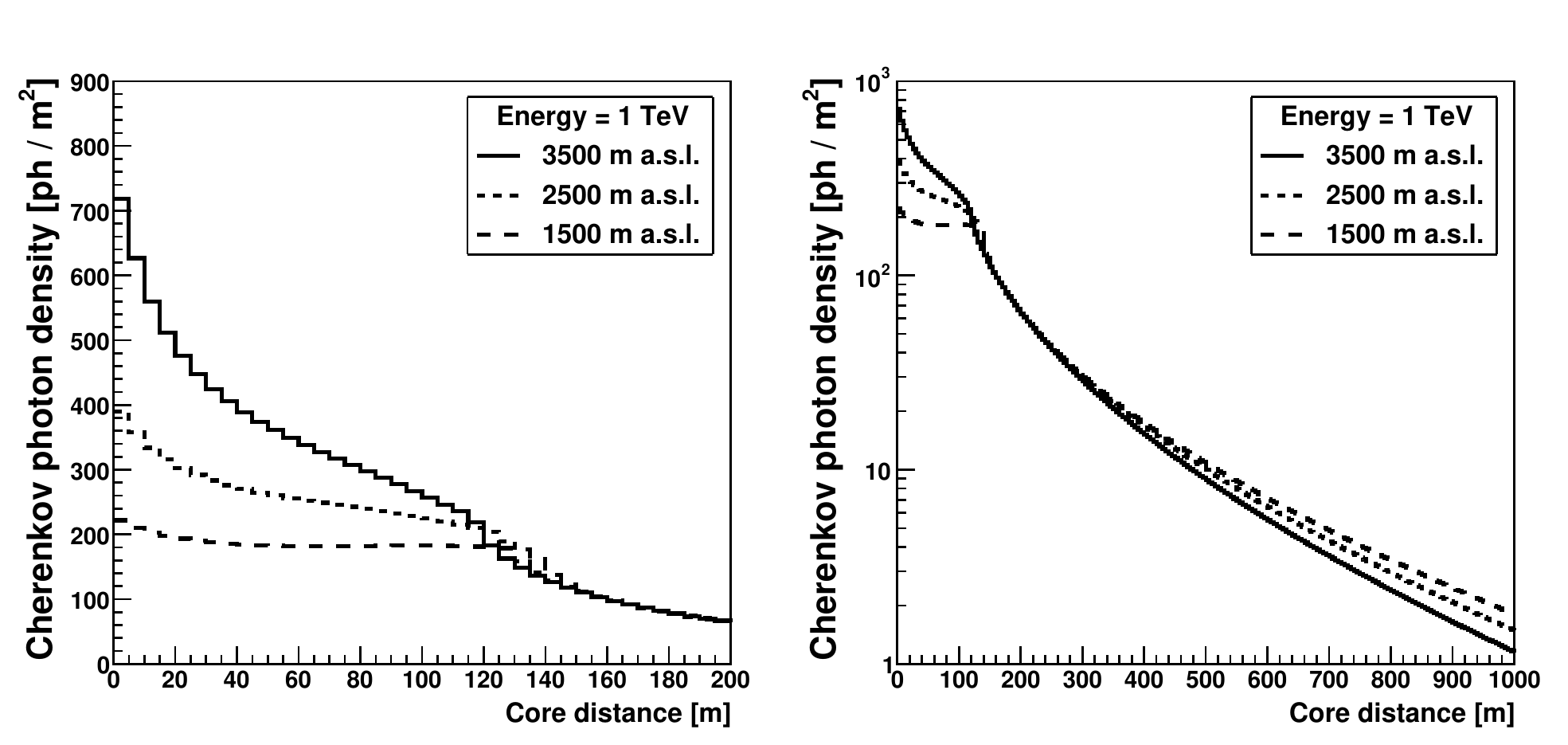}
\centering\includegraphics[width=0.9\linewidth]{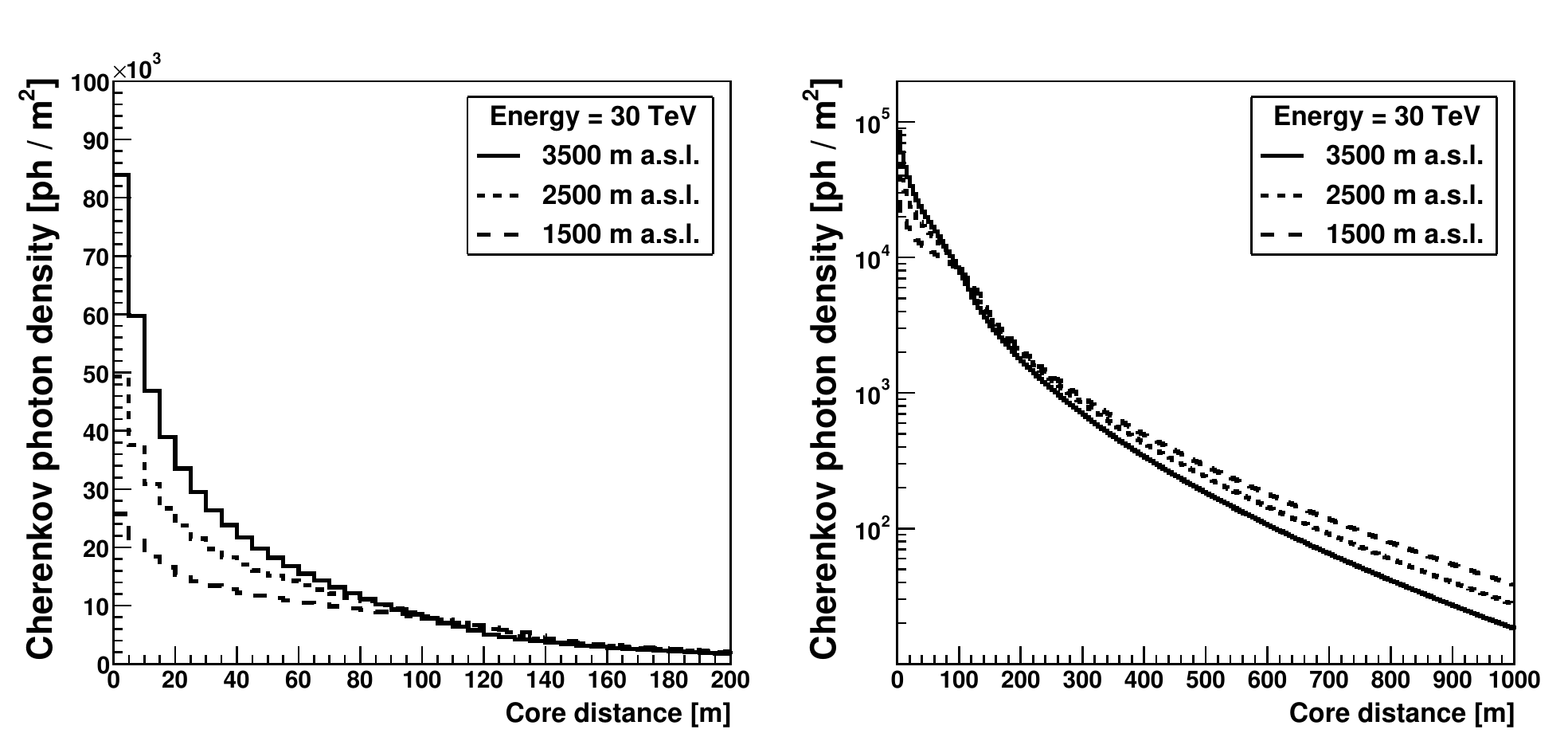}

\caption{Average Cherenkov light lateral distributions (for wavelengths between 300 and 600 nm), at ground levels from 1500 m (dot-dashed lines) to 3500 m (solid lines) above sea level, produced by vertical gamma-ray showers. Three different photon energies are shown: 30 GeV (top), 1 TeV (middle) and 30 TeV (bottom). \textit{Left}) Cherenkov photon density in linear scale close to the core position. \textit{Right}) Broader core distance ranges and logarithmic scale on the Cherenkov photon density. The geomagnetic field used corresponds to H.E.S.S. site (Namibia). No atmospheric absorption is considered.}
\label{fig:lateral}
\end{center}
\end{figure}

The operational altitude of IACTs sets the average stage of development in which EASs are measured \cite{impactAtmospherics}. Therefore the altitude of the IACTs influences the quality of the measurements in several ways:

%There are several effects associated with the altitude of the observatory, mainly related to the distance between the shower maximum and IACTs:

\begin{itemize}

\item for a given gamma-ray energy, the intensity of Cherenkov light close to the shower axis (less than $\sim$150 m) increases at higher altitudes (see Fig. \ref{fig:lateral}, left panels)

\item for gamma-rays with energy above $\sim$200 GeV, Cherenkov photon density at large core distances is reduced at higher observational altitudes (see Fig. \ref{fig:lateral}, right panels)

\item for a given impact parameter\footnote{Distance projected on ground between the center of the Cherenkov light pool and the IACT}, the centroid (i.e. center of gravity) of shower images will be shifted towards the camera edge for higher altitude sites. These images get truncated due to the limited field of view of each telescope, therefore limiting the shower distance accessible range

\item the contribution from charged particles penetrating to ground level increments the fluctuations of gamma-ray images detected by IACTs close to the shower axis. These fluctuations increase the variance of the shape and total charge of shower images, decreasing background rejection efficiency. This effect increases with altitude

\end{itemize}

These effects are translated to lower energy thresholds for higher construction altitudes and reduced performance at energies above $\sim200$ GeV. Considering the CTA sub-systems individually \cite{APP_CTA_MC,Hassan-2015}: 

\begin{itemize}

\item LSTs: at higher altitude sites, more Cherenkov photons reach the telescopes at ground level (see Fig. \ref{fig:lateral}, top panel), providing a lower threshold energy, although angular and energy resolution may be degraded due to the presence of charged particles close to the ground.

\item MSTs: telescope spacing is comparable to the crossover point of the Cherenkov light lateral distributions at different altitudes (see Fig. \ref{fig:lateral}, right panels), so modest performance differences are expected for intermediate energies (200 GeV to 5 TeV). Higher altitude sites reduce telescope multiplicity (i.e. number of telescope images obtained for each shower) but increase the intensity of the recorded shower images.

\item SSTs: telescope multiplicity will be reduced at high altitudes due to the sum of two effects: a reduced atmospheric volume is found within the optical field-of-view of the telescopes (producing bigger images that may be truncated within the IACT camera) and the lower Cherenkov light density emitted by EASs at large impact distances (see Fig. \ref{fig:lateral}, right panels).

\end{itemize}

As shown in Table \ref{table:sites}, there are large differences in altitudes of CTA candidate sites, ranging from 1640 m (Aar, Namibia) to 3600 m (San Antonio de los Cobres, Argentina). 

\subsection{Geomagnetic field intensity}
\label{subsec:gm}

Along the development of EASs, the Earth's magnetic field exerts Lorentz forces on the generated charged particles bending their trajectories. These forces produce a larger lateral spread on charged particles with low energies, mainly $e^{-}/e^{+}$ pairs generated in the EAS, producing a broadening effect on the lateral development of air showers not negligible compared with Coulomb scattering \cite{concconi:1953}, leading to a distortion of the Cherenkov light pool and the shower image shape.%\todo{Add a figure?}

The magnetic force depends on the angle formed between the trajectory of the charged particle and the \gls{gf}. In first approximation, the direction of the trajectories can be approximated by the one of the shower axis and the force is perpendicular to both the \gls{gf} and the axis. The Lorentz force intensity:
\begin{equation}
F_L=q(\vec{v}\times \vec{B}) \propto q\vec{B}_{\bot}
\label{eq:lorentz_}
\end{equation}
is proportional to the perpendicular component of the GF ($\vec{B}_{\bot}$) with respect to the shower direction  (with $B_y = 0$, see \cite{Reyes:2009}):
\begin{equation}
\vec{B}_{\bot} = B_z\sin \theta \sin \phi ~\vec{i} + (B_z \cos \theta - B_z \sin \theta \cos \phi) ~\vec{j} + B_x \sin \theta \sin \phi ~\vec{k}
\label{eq:lorentz}
\end{equation} 
where $\theta$ and $\phi$ are respectively the zenith and azimuth angles of the shower axis and $B_x$ and $B_z$ are the local horizontal and vertical \gls{gf} intensity respectively (H and Z, aligned with $\vec{i}$ and $\vec{k}$), fixed by the location of the observatory. The ($\vec{i}$,$\vec{j}$,$\vec{k}$) correspond to the CORSIKA coordinate frame \cite{corsika}, and point to the magnetic North, West and Zenith (downwards) directions, respectively. Eq. \ref{eq:lorentz_} and \ref{eq:lorentz} show a direct dependency between the Lorentz effect on $e^\pm$ and the direction of the particles. This effect disrupts the cylindrical symmetry of pure electromagnetic cascades, broadening the shower development and the Cherenkov light pool along the East-West direction, distorting the shape of recorded images \cite{Szanecki_2013, ICRC_2008_GF}. In general, higher \gls{gf} intensities slightly increase energy thresholds and degrade angular and energy resolution at low energies.

Table \ref{table:sites} shows the horizontal and vertical components of the intensity of the GF at the different CTA candidate sites. The GF intensity is similar at northern sites, but significantly lower in South America as compared with Southern Africa due to the South Atlantic Anomaly (SAA). However, the horizontal component of the GF, the most relevant parameter for observations near zenith, is lower in South Africa compensating GF differences between Southern Hemisphere sites. Given these considerations, no major performance differences among sites in a given Hemisphere are expected due to the effect of the GF.

\subsection{Night-sky background}
\label{subsec:nsb}

The \gls{nsb} is the diffuse light from the night sky and accounts for the visible light coming from several sources. In the case of IACTs (sensitive to 300--600 nm photons), the main contribution comes from, in order of decreasing contribution, the airglow, zodiacal light and starlight scattered by interstellar dust. Other sources may become dominant under certain conditions, such as Moon light or anthropogenic light. These photons enter the telescope optics producing accidental triggers and increasing noise in the images. Accidental triggers are easily suppressed within the standard IACTs analysis (rarely mimicking low energy showers), but affect the data acquisition performance decreasing the acceptance (due to both the associated dead time following accidental triggers, and the required increased trigger thresholds). Increased \gls{nsb} levels will, in general, increase energy thresholds resulting in a reduction of performance with respect to dark sky observations.

CTA observations up to 5 times the NSB level found in dark sky patches away from the galactic disk are anticipated when the moon is above the horizon. The natural \gls{nsb} levels observed at each site (measured using an Autonomous Tool for Measuring Observatory Site COnditions PrEcisely (ATMOSCOPE), \cite{sitePaper3}) increase with altitude as a result of the reduced attenuation of zodiacal and starlight. The NSB of all candidate sites was measured \cite{NSBmarkus} and results showed similar levels, with the exception of Teide (Tenerife, Spain), where the level of \gls{nsb} is 30\% higher with respect to other sites at similar altitude after subtraction of starlight, due to anthropogenic light from nearby cities. Note that the candidate site in La Palma, essentially identical to the Tenerife site in all other respects, has a smaller NSB, similar to the rest of candidate sites.

\section{Monte Carlo Simulations of Site Candidates for CTA}

The evaluation studies presented in the following uses detailed Monte Carlo (MC) simulations of the instrument in development. As presented in \cite{APP_CTA_MC, Hassan-2015}, these are performed by defining a large telescope layout, comprising few hundreds of telescopes of different types distributed over an area of around 6 km$^2$. From this master layout, subsets of telescopes are selected and analysed as feasible CTA layouts. Simulations of EASs initiated by gamma rays, cosmic-ray nuclei and electrons are generated using an EAS simulation software, CORSIKA \cite{corsika} together with the simulation of the telescope response using \tmtexttt{sim\_telarray} \cite{Konrad:2008}, software packages extensively used and validated by the High Energy Gamma Ray Astronomy (HEGRA) and High Energy Stereoscopic System (H.E.S.S.) experiments. 

The EASs are simulated independently for each of the candidate sites, with specific atmospheric density profiles, altitudes and GFs (direction and intensity). Site atmospheric models (density and refraction index as a function of the altitude) were generated using the NRLMSISE-00 model \cite{atmModel} and cross-checked against radiosonde data, where available near the sites. A total of 3 Northern and 6 Southern Hemisphere sites were simulated, all listed in Table \ref{table:sites}, together with their altitudes and geomagnetic field strengths (\cite{sitePaper1, sitePaper2}). 

Closely located site candidates: Yavapai and Meteor Crater in the US, Armazones and Paranal in Chile and Tenerife and La Palma in Spain have similar characteristics. Therefore, only one site has been simulated in each case. Note ``Aar@500m" is a hypothetical site located at Aar, Namibia with an assumed altitude of 500 m, computed to evaluate CTA performance at a significantly lower altitude.

Simulated showers include gamma rays (from a point source) and background (mainly protons and electrons), with protons ($\sim$ 100 billion events per site) being the particle type consuming most of the CPU time and disk space resources, even though few of them trigger and pass the selection cuts. While nuclei (helium through iron) account for about a quarter of the showers triggering a telescope system, they are easily distinguished from gamma-ray showers in the analysis (see section \ref{sec:analysis}), not contributing significantly to the background after cuts \cite{APP_CTA_MC}. As a consequence, simulation of showers induced by nuclei was carried out only for a few selected sites. A minimum of 2 triggered telescopes were required for each shower to be stored. Most of the simulations were produced for zenith angles of 20 deg (except for 3 sites, for which simulations were also done at 40 deg).

\begin{table}[htp]

\begin{center}
\begin{tabular}{l|c|c|c|c}
Candidate site name & Lat., Long. & Altitude & \textit{B}$_{\mathrm{x}}$ & \textit{B}$_{\mathrm{z}}$ \\
       &                       [deg]         &  [m]   & [$\mu$T] & [$\mu$T]\\
\hline\hline
Aar (Namibia) & $26.69$ S $6.44$ E & 1640 & 10.9 & -24.9 \\ %ATM24
Armazones (Chile) & $24.58$ S $70.24$ W & 2100 & 21.4 & -8.9 \\ %ATM26
Leoncito@2640 m (Argentina)              & $31.72$ S $69.27$ W & 2640 & 19.9  & -12.6 \\
Leoncito@1650 m (Argentina)     & $31.41$ S $69.49$ W& 1650 &19.9 & -12.6 \\
\begin{tabular}{@{}c@{}}San Antonio de los Cobres  \\ (SAC;Argentina)\end{tabular}       &  $24.05$ S $66.24$ W   & 3600 & 20.9 & -8.9  \\
\hline
Meteor Crater (USA)  & $35.04$ N $111.03$ W & 1680 & 23.6 &  42.7 \\
San Pedro Martir (SPM; Mexico)  & $31.01$ N $115.48$ W & 2400 & 25.3 & 38.4 \\
Teide, Tenerife (Spain)  & $28.28$ N $16.54$ W & 2290 & 30.8 & 23.2 \\
\hline
Aar@500 m (hypothetical site) & $26.69$ S $6.44$ E & 500 & 10.9 & -24.9 \\
\end{tabular}
\end{center}
\caption{\label{table:sites} Summary table of all simulated CTA candidates sites.
The strength of the geomagnetic field is given by its horizontal (\textit{B}$_{x}$) 
and downwards pointing (\textit{B}$_{\mathrm{z}}$) components (see section \ref{subsec:gm}). Meteor Crater and Tenerife simulations represent also the nearby sites of Yavapai (Arizona, USA) and Roque de los Muchachos Observatory (La Palma, Spain) respectively.
}
\end{table}

To account for the effect of the geomagnetic field for different azimuth angles, simulations were carried out with showers coming from both the north and south directions. The assumed \gls{nsb} level corresponds to dark-sky observations towards an extra-galactic field at each site. Telescope optical system and hardware settings and available observation time are assumed to be identical at all sites \cite{APP_CTA_MC, Konrad:2008}. At each site, individual telescope trigger thresholds are set so that the rate of accidental triggers is equal to the rate expected from cosmic rays. An additional lower-scale production was carried out with elevated NSB levels (by 30\% and 50\%, just for ``Leoncito@2640" site) to estimate the impact of increased NSB levels on CTA performance (see Sec. \ref{subsec:nsb}).

A total of 229 telescope positions were simulated for each Southern Hemisphere site, with up to 7 different telescope types \cite{Hassan-2015} (with many positions used by several telescopes): LST, two MST\footnote{Modified Davies-Cotton (DC) \cite{MSTgamma} and Schwarzschild-Couder (SC) designs \cite{MSTgamma,Prod2_SCMST}.}, and up to four variants of SST\footnote{Two designs of SC-SSTs, the ASTRI (\textit{Astrofisica con Specchi a Tecnologia Replicante Italiana}) and the GCT (Gamma-ray Cherenkov Telescope), both with primary mirror diameters of 4 m, and 2 DC-SSTs, the SST-1M and 7m-SST with a single 4 m and 7 m diameter mirror respectively \cite{Montaruli:2015}.}. 

\begin{figure}[htbp]
\begin{center}
\centering\includegraphics[width=1\linewidth]{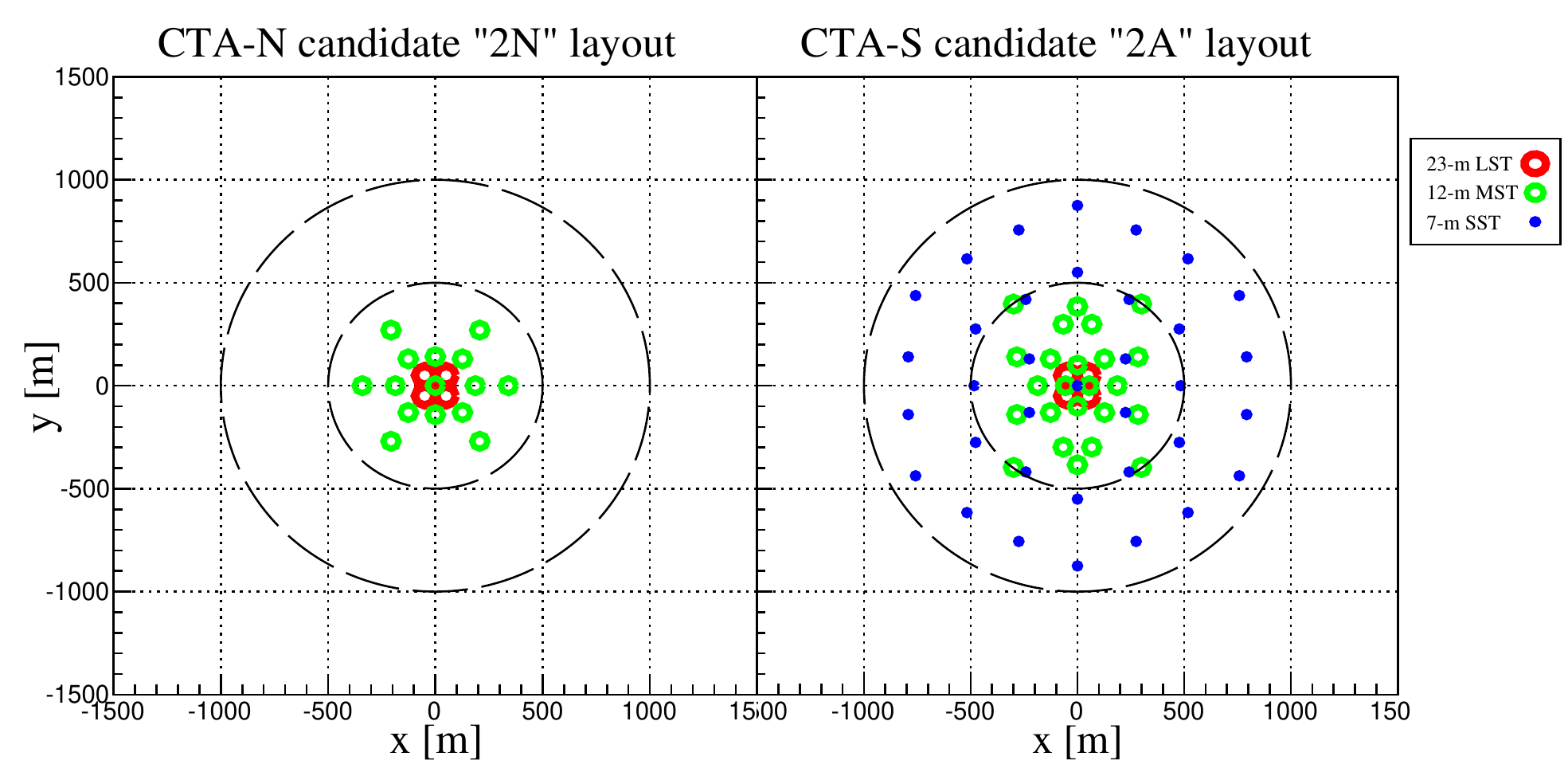}
\caption{Considered array layouts for CTA North (left; named ``2N") and CTA South (right; named ``2A"). The colored markers indicate the position of the corresponding telescopes on the ground. Red circles: Large-Sized Telescopes; Green circles: Medium-Sized Telescopes; Blue circles: 7-m class Small-Sized Telescopes.}
\label{Fig:Arrays}
\end{center}
\end{figure}

The array layout considered  here consists of 4 LSTs, 24 DC-MSTs, and 35 7m-SSTs (see Figure \ref{Fig:Arrays}, Right) for the Southern sites, and of 4 LSTs and 15 DC-MSTs (see Figure  \ref{Fig:Arrays}, left) for the layout for the Northern sites. The array layouts and individual telescope characteristics used in the present work (see Table \ref{table:telescopes}) are not identical to those in the final CTA design, but the differences are not expected to be relevant for the purpose of comparing the sites.

\begin{table}[htp]

\begin{center}
\begin{tabular}{l|c|c|c}
 & LST & MST & DC-SST \\
\hline\hline
CTA-N layout [telescopes] & 4 & 15 & 0 \\
CTA-S layout [telescopes] & 4 & 24 & 35 \\
\hline
No. of mirrors & 198 & 84 & 120 \\
Mirror tile diameter [m] & 1.51 & 1.20 & 0.60 \\
% Mirror gaps [m] & 0.03 & 0.02 & 0.02 \\
Mirror dish area [m2] & 386.9 & 103.9 & 37.2 \\
% Outer diameter [m] & 24.25 & 12.54 & 7.48 \\
Mean edge diameter [m] & 23.21 & 12.10 & 7.19 \\
Telescope focal length [m] & 28.0 & 16.0 & 11.2 \\
Mirror facet focal length(s) [m] & var. & 16.07 & 11.2 \\
Radius of curv. (at dish center) [m] & 56.0 & 19.2 & 11.2 \\
Central hole diameter [m] & 1.57 & 1.24 & 0.64 \\
\hline
Camera pixels & 1855 & 1855 & 1296 \\
Pixel size [mm] & 50 & 50 & 50 \\
Field of view [deg] & 4.6 & 8.1 & 9.1 \\
\end{tabular}
\end{center}
\caption{\label{table:telescopes} Summary table of some relevant parameters used within \tmtexttt{sim\_telarray} telescope simulation. The mirror facet focal length of LSTs are variable, adjusted to the parabolic shape of the dish. Note these parameters, especially for the DC-SST, do not correspond to the current specifications of CTA telescopes.}
\end{table}

All results shown in the following sections refer to point-like gamma-ray sources located at the centre of the field of view and observed at a zenith angle of 20 deg. Results are averaged between two different azimuth directions (all telescopes pointing towards the north or the south), unless otherwise stated. A typical MC set for one site comprises about a billion simulated gamma-ray and electron, and about 100 billion proton showers. The simulation requires substantial computing resources: the simulation of a single candidate site requires between 10--20 million HEP-SPEC06\footnote{The HEP-wide benchmark for measuring CPU performance} CPU hours and $\sim$ 100 TB of event data are written to disk. A large fraction of the MC production used the European Grid Infrastructure (EGI), utilising the DIRAC framework as interware \cite{dirac-general, Arrabito-2015}. Simulations were carried out on the CTA computing grid as well as on the computer clusters of Max-Planck-Institut f\"ur Kernphysik (MPIK).

\section{Analysis}
\label{sec:analysis}

To process the MC production and evaluate the performance of each candidate site, several independent analysis tools derived from packages belonging to different IACT experiments have been used: baseline analysis from H.E.S.S., \textit{Eventdisplay} from VERITAS, \gls{mars} and FAst Simulation for imaging air Cherenkov Telescopes (FAST) (details can be found in \cite{APP_CTA_MC, evnDisplay, MARS, Wood:2014}). Although each analysis chain utilises techniques with subtle differences, all of them consist of the following basic steps:

\begin{itemize}

\item Waveform integration: Each pixel charge and signal arrival time are calculated from the pixel charge time evolution (as simulated by \tmtexttt{sim\_telarray}) for each triggered telescope. %This is performed in a two-pass integration process \cite{Holder-2006} using both charge and timing information to find the optimal placement of the trace integration window.

\item Image cleaning and parametrisation: image cleaning algorithms are applied to separate pixels likely illuminated by Cherenkov photons from those just containing noise or NSB photons. Substantially different methods have been used by alternative analysis chains, such as a 2-level next-neighbour  \cite{Daum:1997} or an aperture cleaning \cite{Wood:2014} approach. The resulting cleaned shower images are then parametrised by the second-moment Hillas analysis \cite{hillas}.

\item Stereoscopic reconstruction: Using the information gathered by all triggered telescopes observing an event, a stereo reconstruction is performed by using the major axes of the cleaned images to reconstruct the direction of the primary particle. Multivariate event classification algorithms (e.g. random forest \cite{Breiman2001}) are used for the energy estimation and background suppression, discerning between cosmic and gamma-ray initiated showers. These algorithms are trained for the two simulated pointing directions separately, for each site candidate, with independent data samples, not used for the performance evaluation.

\item Performance estimation: Optimal cuts are determined in order to calculate the resulting performance, expressed by the Instrument Response Function (IRF)\footnote{The IRF relates the source-emitted photons with the detected events, allowing the computation of gamma-ray fluxes as a function of time, energy and direction.}. Differential sensitivity is maximised for each energy bin by optimising the cuts on the shower arrival direction, \textit{hadronness}\footnote{The \textit{hadronness} variable, defined between 0 and 1, indicated how likely is that the shower has hadronic origin \cite{hadronness}.} (or equivalent) and minimum telescope multiplicity (larger than 1). Similarly as in \cite{APP_CTA_MC}, sensitivity is computed by requiring five standard deviations (5$\sigma$) for a detection at each energy bin (equation 17 from \cite{LiMa}). A ratio of the off-source to on-source exposure of 5 is considered, a plausible value assuming the amount of reflected regions that will be accessible to CTA. In addition, the signal excess is required to be larger than 10, and at least five times the expected systematic uncertainty in the background estimation (1\%). Unless otherwise stated, all differential sensitivities shown in this work are calculated for 50 hours of observation time.

\end{itemize}

The analysis chains used show consistent differential sensitivity \cite{Hassan-2015}, considering the significant differences between them (image cleaning algorithms, shower reconstruction, quality cuts and background rejection power). In addition, the conclusions of this work do not change with the selected analysis chain to perform the different performance comparisons. It is expected that the performance of the future CTA reconstruction pipeline with more sophisticated analysis chains (e.g. improved stereo reconstruction \cite{MAGIC_disp, HESS_disp, VERITAS_disp}, image cleaning \cite{Maxim_cleaning_magic, Maxim_cleaning_cta} or \textit{model analysis} \cite{modelAnalysis}) will provide a significant improvement as compared to the results presented in the following, as they are obtained with traditional analyses optimised for the current generation of IACTs, with 2--5 telescopes in operation. 

CTA candidate sites scientific performance evaluation was carried out using all available analysis chains cited in this section. From here on, for clarity, results shown correspond to the \textit{Eventdisplay} analysis.

% \begin{figure}
% \begin{center}
% \centering\includegraphics[width=1.0\linewidth]{compareAnalysisSensitivities.pdf}
% \caption{Differential sensitivity for the CTA-S ``2Q" candidate array (50 hours of observation on-axis of a point-like source, N/S pointing average) calculated with 3 alternative analysis chains: \textit{Black}: MARS analysis. \textit{Red}: \textit{Baseline} analysis. \textit{Green}: \textit{evnDisplay} analysis. The ``2Q" layout is a variant of ``2A", using a more realistic 4 m SST variant (see \cite{Hassan-2015}).}
% \label{fig:analysis_compare_2Q}
% \end{center}
% \end{figure}

\section{Science Performance}
\label{sec:sciencePerformance}

\begin{figure}
\begin{center}
\centering\includegraphics[width=1.0\linewidth]{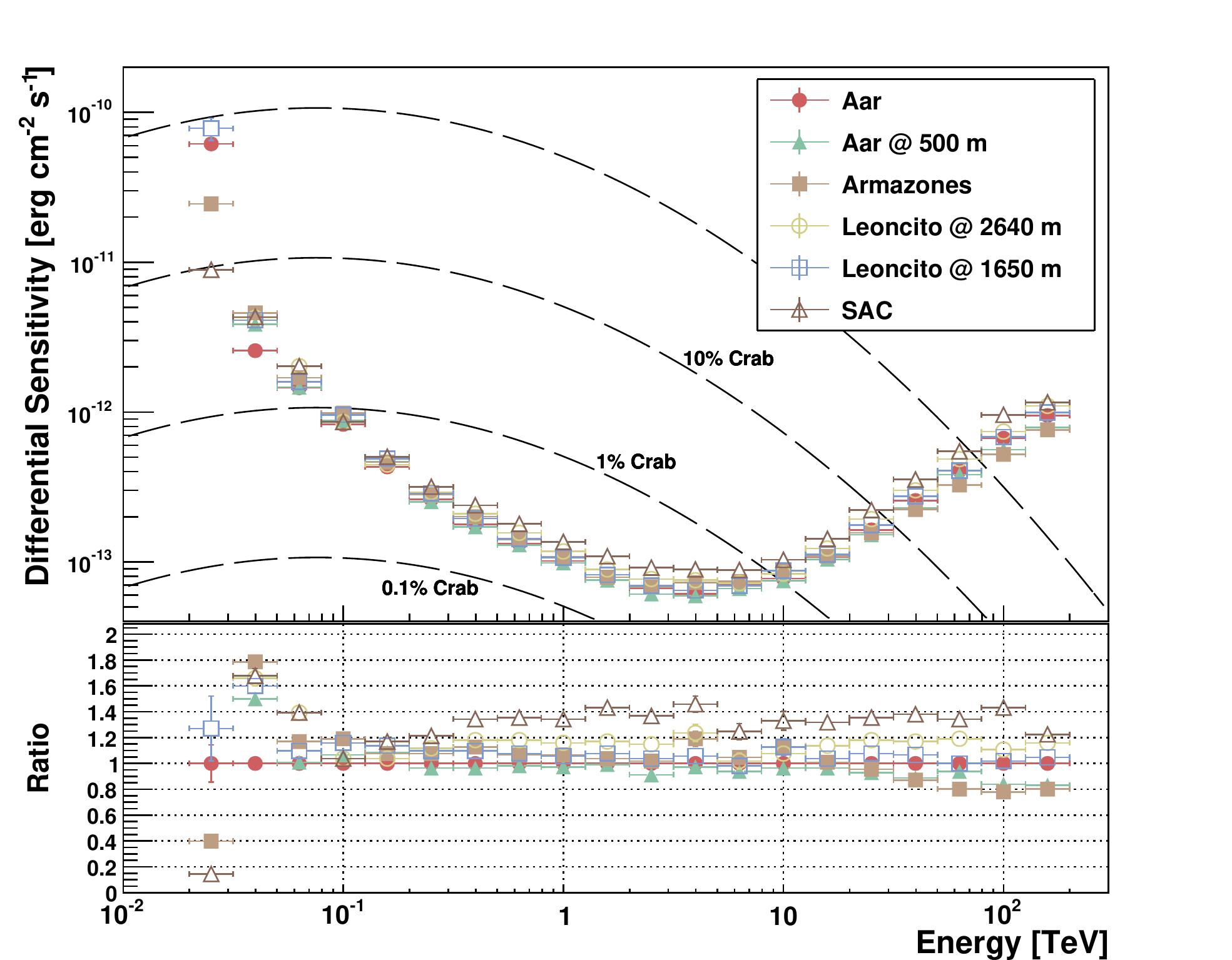}
\caption{On-axis differential point-source sensitivity for the considered CTA site candidates located in the Southern Hemisphere (see Table \ref{table:sites}) as a function of the reconstructed energy. Average sensitivities calculated from telescopes pointing towards the North and the South at 20 deg in zenith angle are shown. The layout candidates ``2A" have been used. Differential sensitivities are derived for 50 h of observations. The dashed lines indicate the flux of a Crab Nebula-like source scaled by the factors indicated in the figure. Horizontal ``error" bars indicate the bin size in energy, while vertical ones show the uncertainty of the flux sensitivity, derived from propagating the statistical uncertainties associated to MC event statistics. Bottom: sensitivity ratios are calculated with respect to the ``Aar" site (smaller ratios mean better sensitivity).}
\label{Fig:DiffSensSouth}
\end{center}
\end{figure}

\begin{figure}
\begin{center}
\centering\includegraphics[width=1.0\linewidth]{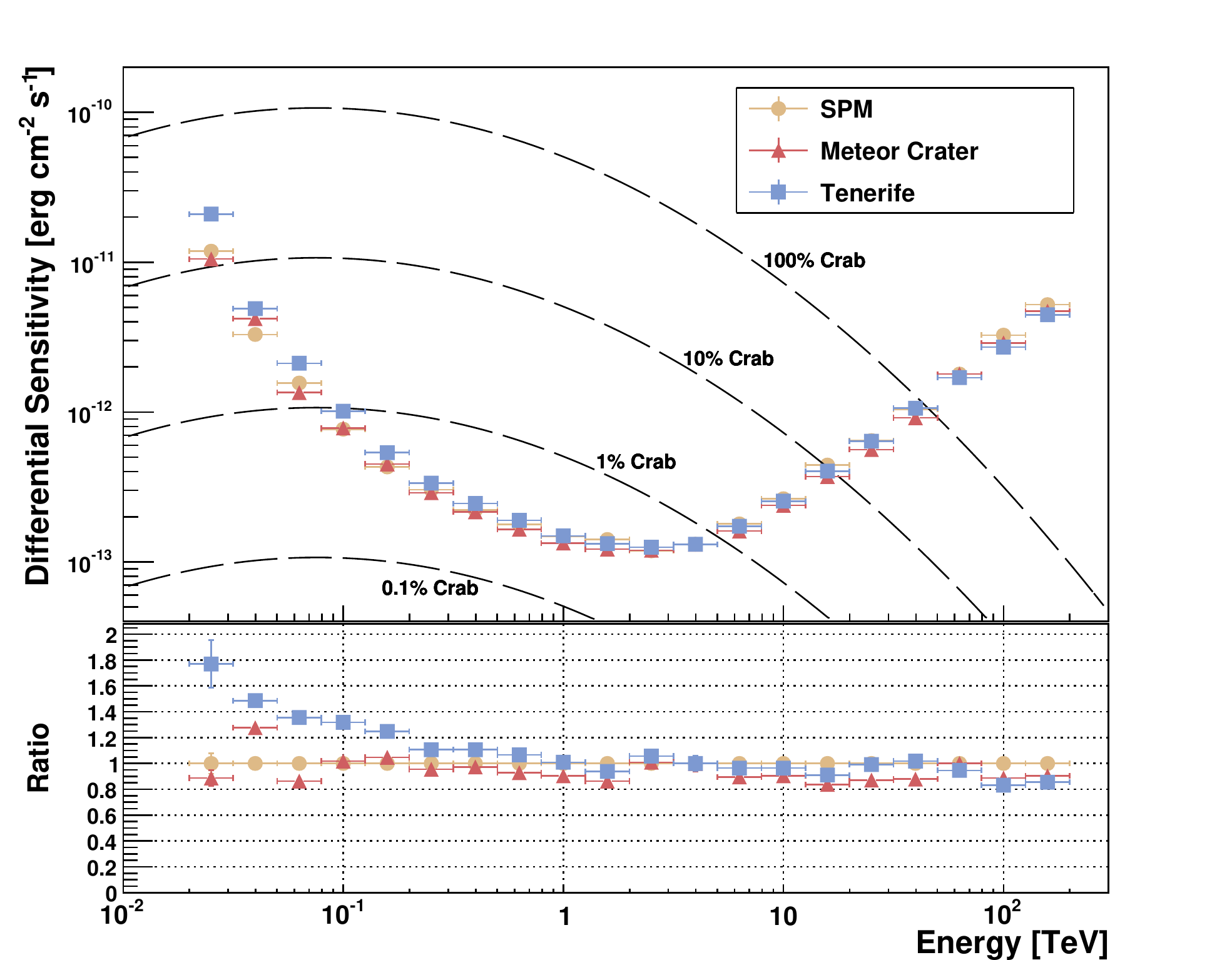}
\caption{Same as fig. \ref{Fig:DiffSensSouth}, but the simulation results for the Northern sites are shown. The layout candidates ``2N" have been used. Bottom: sensitivity ratios are calculated with respect to the ``SPM" site.}
\label{Fig:DiffSensNorth}
\end{center}
\end{figure}

\begin{figure}
\begin{center}
\centering\includegraphics[width=1.0\linewidth]{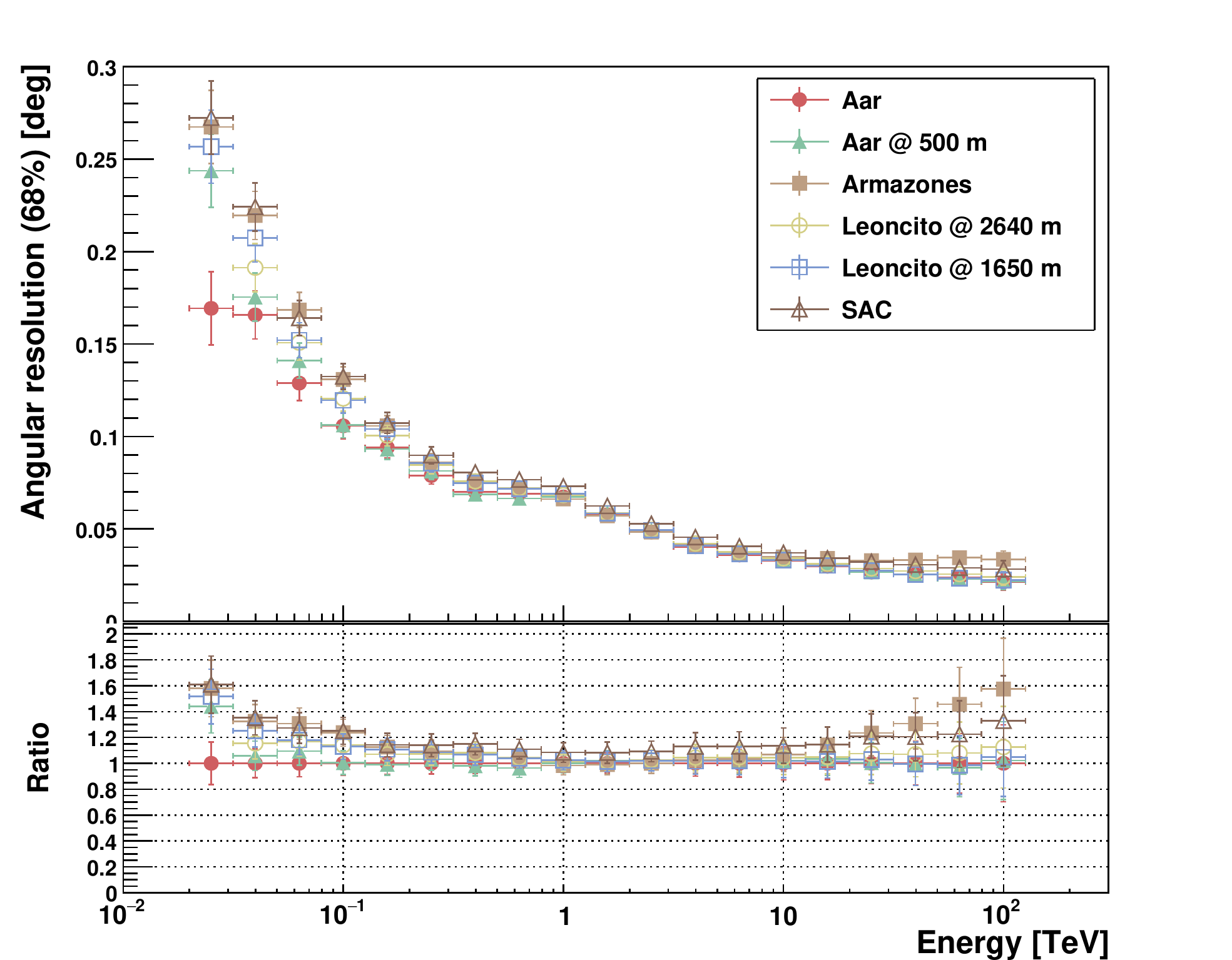}
\caption{Angular resolution for the considered CTA site candidates located in the Southern Hemisphere (see Table \ref{table:sites}) as a function of the reconstructed energy. Angular resolution is defined as the angle within which 68\% of reconstructed gamma-ray events fall (surviving gamma-hadron separation cuts), relative to the true direction. Average resolution from telescopes pointing towards the North and the South at 20 deg in zenith angle are shown. Note that this analysis is not optimised to provide best angular resolution, but rather best point-source sensitivity. Higher resolution is possible at the expense of some collection area. The layout candidates ``2A" have been used. Horizontal ``error" bars indicate the bin size in energy, while vertical ones show the uncertainty of the angular resolution, derived from propagating the statistical uncertainties associated to MC event statistics. Bottom: angular resolution ratios are calculated with respect to the ``Aar" site (smaller ratios mean better resolution).}
\label{Fig:AngResSouth}
\end{center}
\end{figure}

\begin{figure}
\begin{center}
\centering\includegraphics[width=1.0\linewidth]{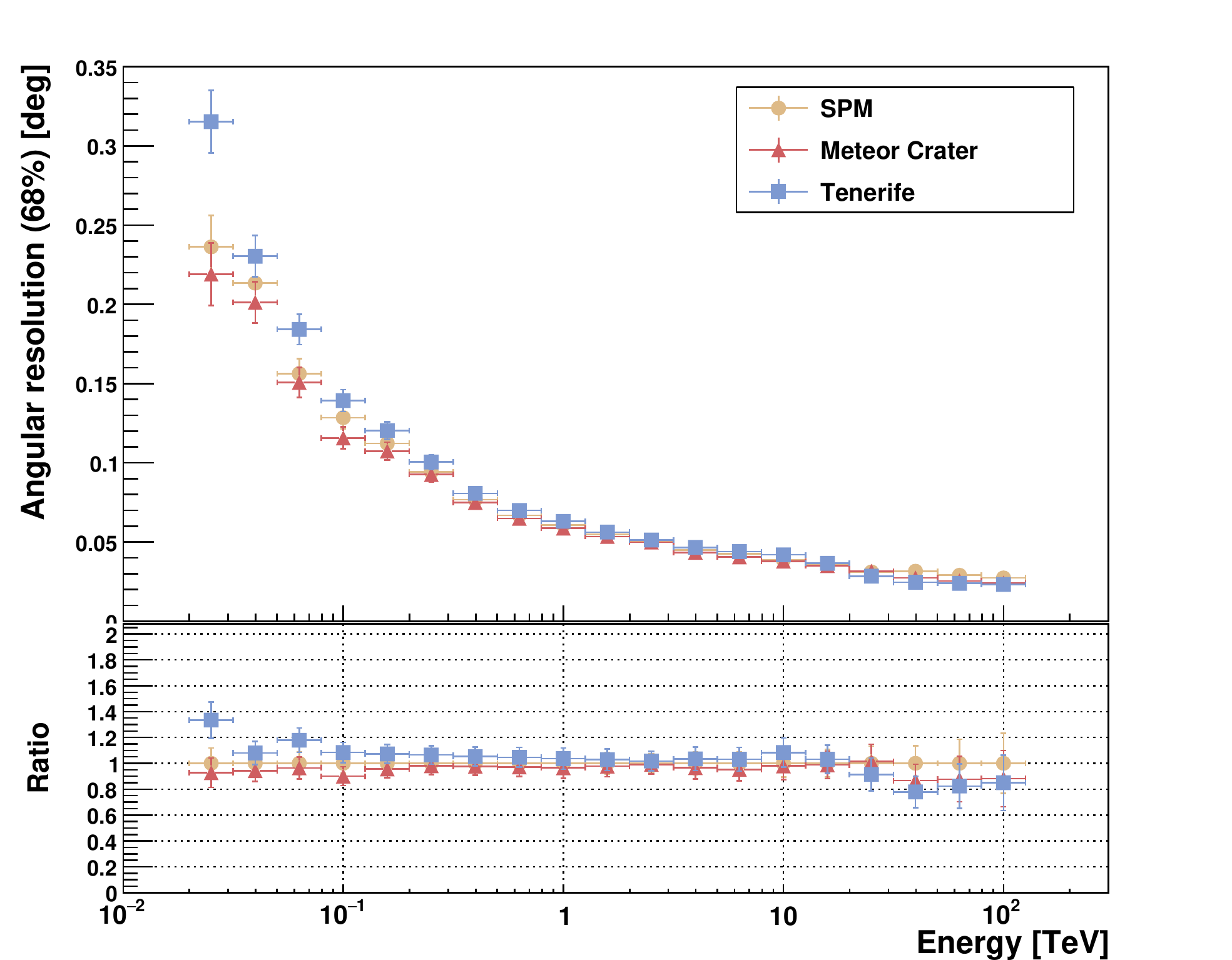}
\caption{Same as fig. \ref{Fig:AngResSouth}, but the simulation results for the Northern sites are shown. The layout candidates ``2N" have been used. Note that this analysis is not optimised to provide best angular resolution, but rather best point-source sensitivity. Higher resolution is possible at the expense of some collection area. Bottom: angular resolution ratios are calculated with respect to the ``SPM" site.}
\label{Fig:AngResNorth}
\end{center}
\end{figure}

\begin{figure}
\begin{center}
\centering\includegraphics[width=1.0\linewidth]{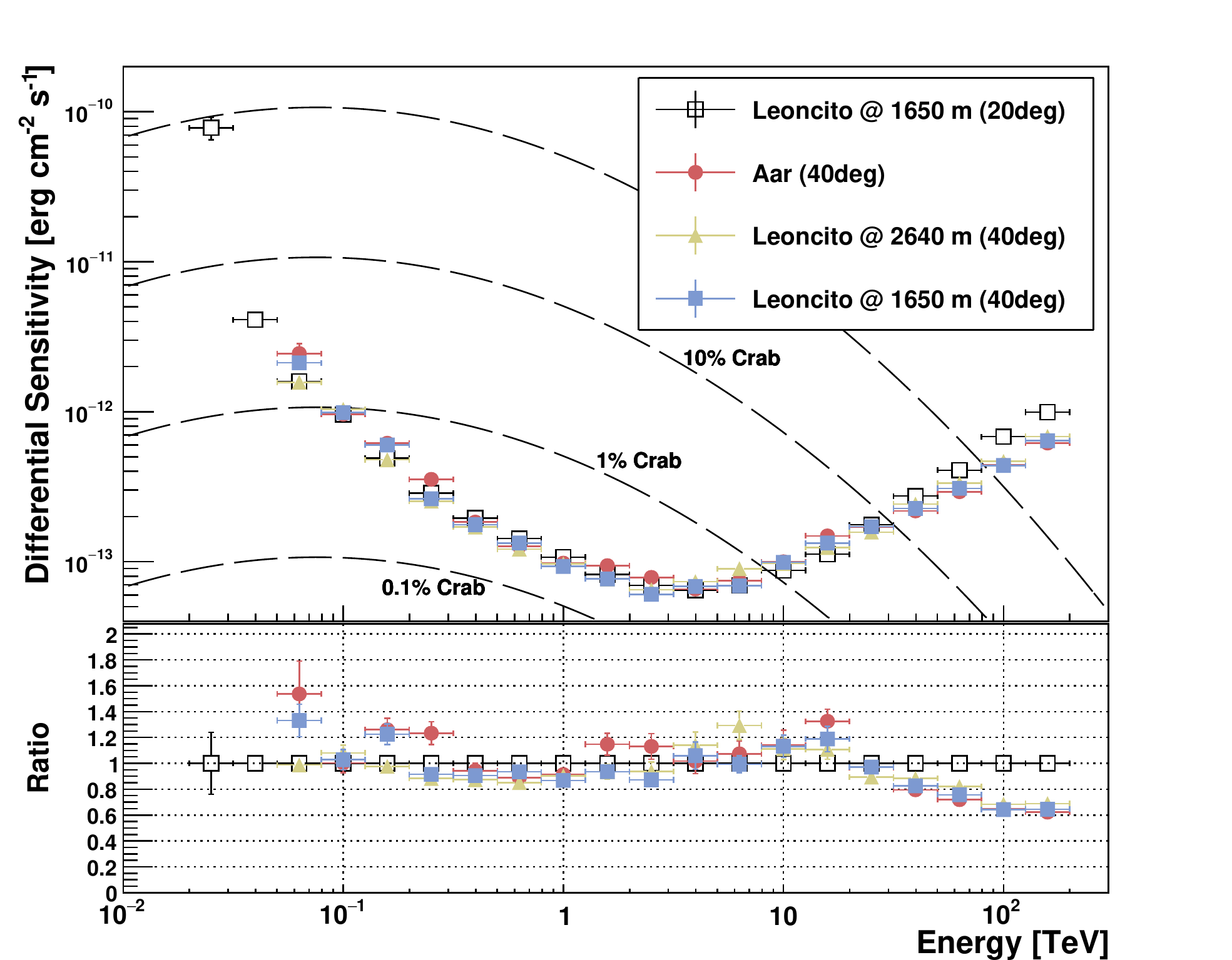}
\caption{On-axis differential point-source sensitivity for CTA-S site candidates pointing at 40 deg zenith angle. Sensitivity of ``Leoncito@1650m" at 20 deg zenith angle is used as reference. Average sensitivities calculated from telescopes pointing towards North and South at 20/40 deg in zenith angle are shown. Differential sensitivities are derived for 50 h of observations.}
\label{Fig:DiffSens40deg}
\end{center}
\end{figure}

As introduced in section \ref{siteParameters}, the primary performance criteria for the site evaluation is the differential sensitivity over the entire energy range of CTA, from 20 GeV to 300 TeV. Five bins of equal logarithmic width are used per energy decade.
 
As good sensitivity is required over the complete energy range defined above, the figure of merit used for the comparison of the science performance at the different site candidates is the so-called performance per unit time (PPUT). It is defined as the geometric mean through individual energy bins of the inverse of the sensitivity normalised to a reference sensitivity:

\begin{equation}
\mathrm{PPUT} =  \left( \prod_{i=1}^{N} \frac{F_\mathrm{sens,ref}(i)}{F_\mathrm{sens}(i)} \right)^{1/N}
\end{equation}

\noindent
where $F_\mathrm{sens,ref}$ is the reference sensitivity and $F_\mathrm{sens}$ the achieved one through $N$ bins in energy, from 30 GeV to 200 (20) TeV for CTA South (North). The reference sensitivity (used for normalization) was derived from the analysis of previous simulations carried out by the Consortium (see \cite{APP_CTA_MC}) for a site at 2000 m altitude and with a geomagnetic field strength and orientation intermediate between that found at the Aar and Tenerife sites. These previous MC simulations for CTA were based on initial and conservative assumptions of telescope parameters and simplified readout systems, therefore PPUT values are expected to be significantly larger than 1, higher for candidates with better (lower) differential sensitivity across the whole energy range.

In order to also evaluate the effect of the GF on the angular resolution over the whole energy range of CTA, a similar figure of merit is defined analog to the PPUT. The Angular Performance (AP) is defined as the geometric mean through individual energy bins of the inverse of the angular resolution normalised by a reference angular resolution:
\begin{equation}
\mathrm{AP} =  \left( \prod_{i=1}^{N} \frac{\Theta_\mathrm{0.68,ref}(i)}{\Theta_\mathrm{0.68}(i)} \right)^{1/N}
\end{equation}
where $\Theta_\mathrm{0.68, ref}$ is the reference angular resolution and $\Theta_\mathrm{0.68}$ the calculated one from each candidate site through $N$ bins in energy, from 30 GeV to 10 TeV (energies in which the effect of the GF is more relevant), defined as the 68\% containment radius (i.e. the angle within which 68\% of reconstructed gamma rays are contained, relative to the simulated direction). Reference angular resolution, similar to the reference sensitivity, was derived from the analysis of a previous production of CTA simulations (see \cite{APP_CTA_MC}). Higher AP will be found for candidates with better (smaller) angular resolution across the whole energy range.

Note the cut optimisation is performed independently for the analysis of each site, maximising differential sensitivity. The angular resolution curves shown in this work are calculated using these cuts, therefore they only represent a conservative estimation of the future angular performance of CTA. Angular resolution improves by imposing tighter cuts (e. g. on event multiplicity) at the expense of reducing differential sensitivity. 
 
\subsection{Performance for dark-sky observations}

The on-axis differential point-source sensitivity and angular resolution as a function of the energy for the considered CTA site candidates is shown in figures \ref{Fig:DiffSensSouth}, \ref{Fig:DiffSensNorth}, \ref{Fig:AngResSouth}, \ref{Fig:AngResNorth} and \ref{Fig:DiffSens40deg}. As shown in these figures, there are significant performance variations between the candidate sites. These curves do not take into account the AAOT differences between candidate sites.

\begin{figure}[htbp]
\begin{center}
\centering\includegraphics[width=0.7\linewidth]{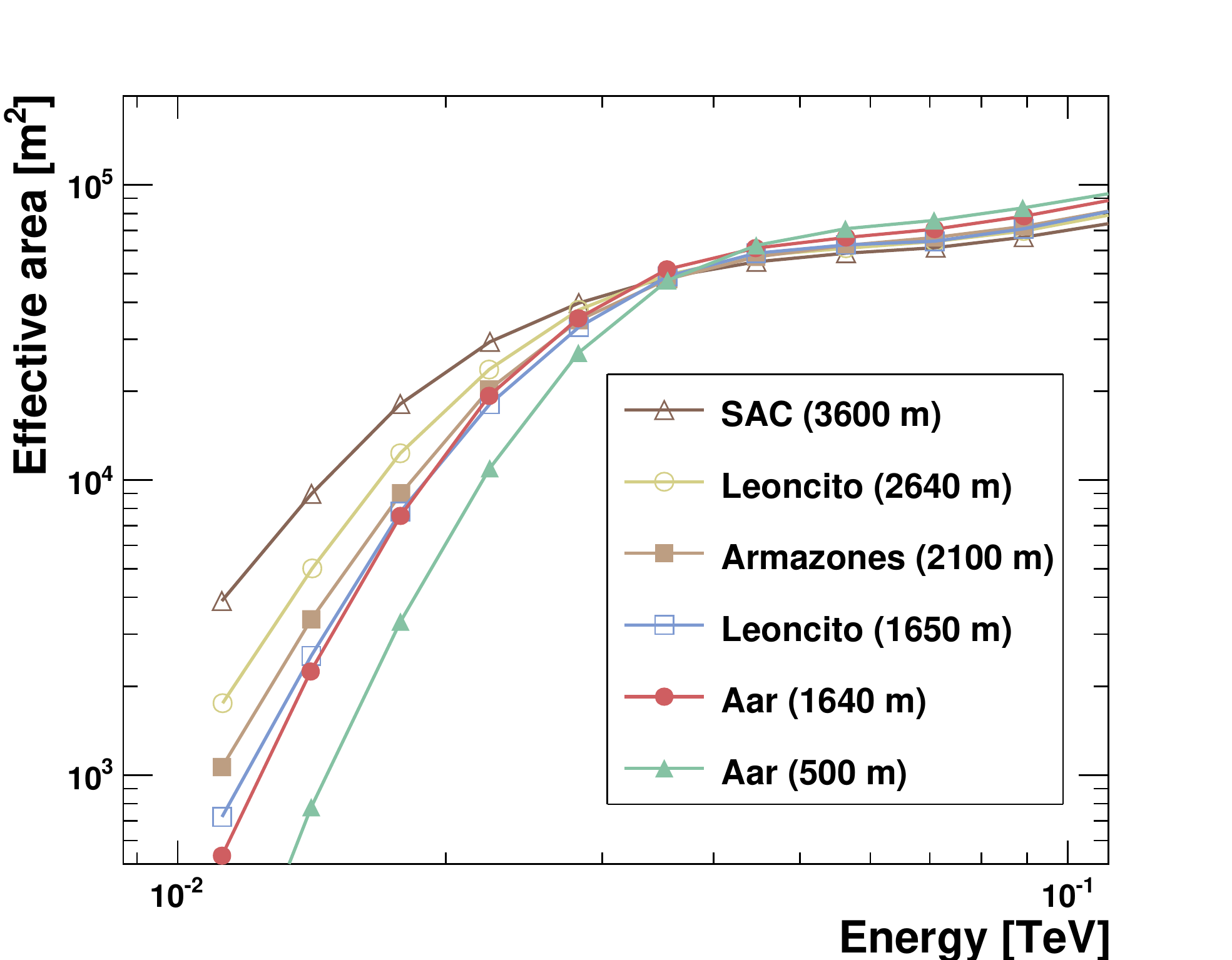}
\caption{Effective area vs true (simulated) energy of several CTA Southern candidate sites, showing how the altitude affects the energy threshold of the observatory. Very loose cuts have been applied in the analysis for this figure, requiring the successful reconstruction of direction and energy only.
}
\label{Fig:EffAreas1}
\end{center}
\end{figure}

Close to the energy threshold of the instrument (E $<$ 50 GeV), the detection is limited by the number of Cherenkov photons hitting the telescopes, and performance differences between sites are dominated by the altitude. As previously described, higher altitude sites are placed closer to the shower maximum, and therefore collect more Cherenkov photons at distances $<$150 m to the shower axis (see Figure \ref{Fig:EffAreas1}). The highest-altitude site, San Antonio de los Cobres (SAC) at 3600 m a.s.l., shows the best performance among all sites for energies bellow 30 GeV. 

\begin{figure}
\begin{center}
\centering\includegraphics[width=0.49\linewidth]{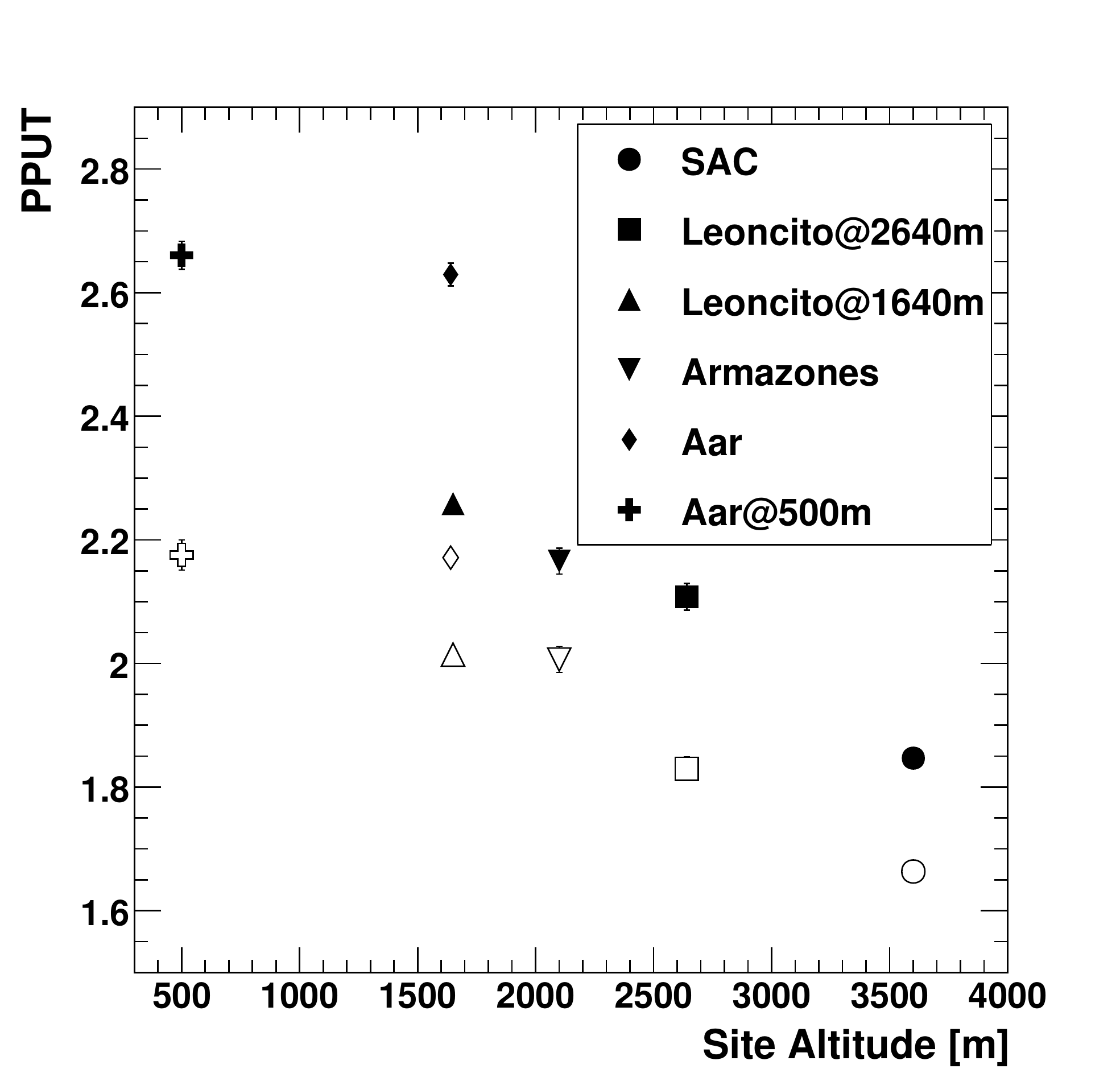}
\centering\includegraphics[width=0.49\linewidth]{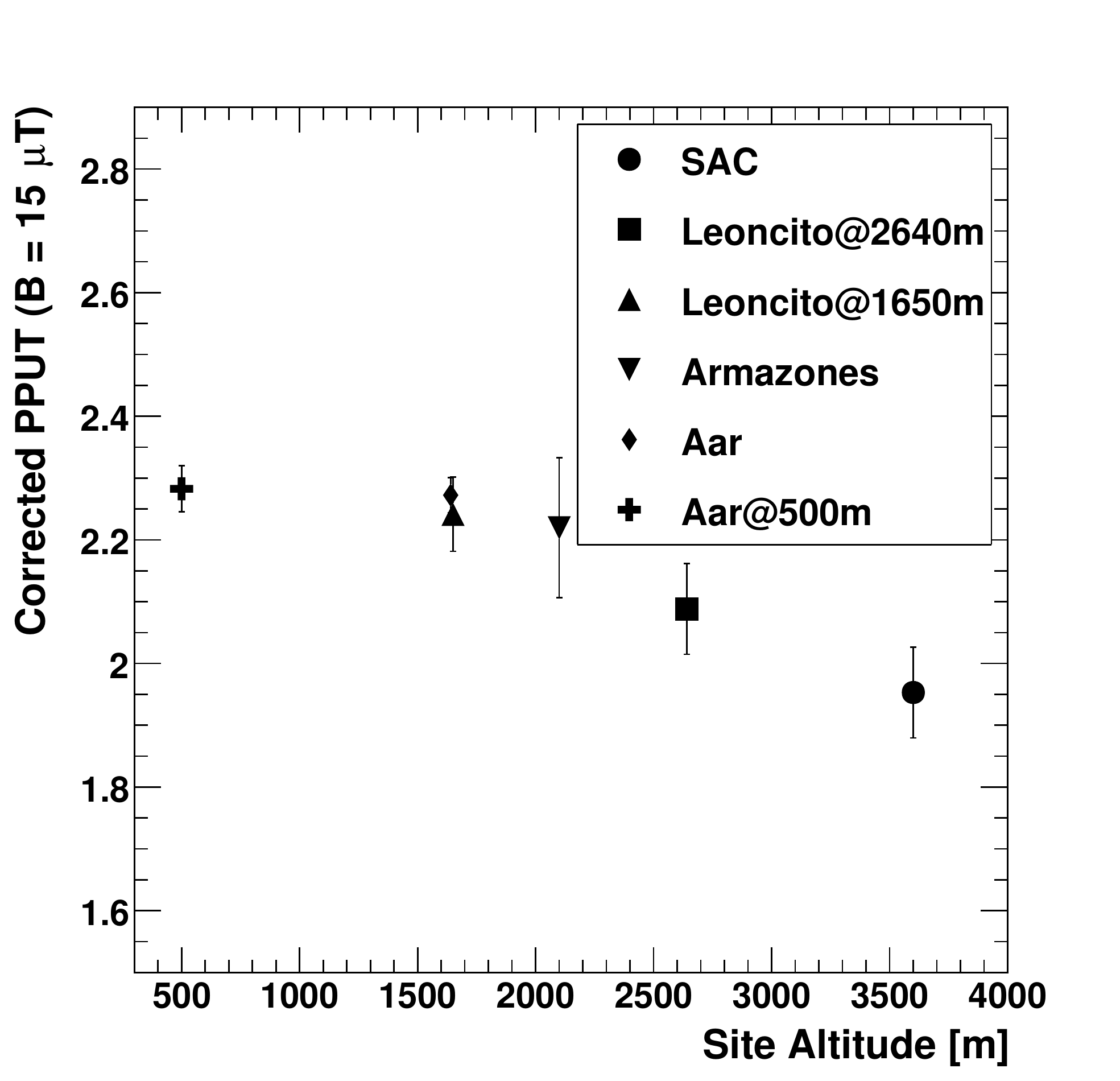}
\caption{Left: PPUT as a function of altitude for the CTA Southern site candidates. Filled (hollow) markers indicate results for arrays with telescopes pointing towards the North (South). Right: Corrected PPUT values as a function of altitude. All sites are corrected to a B = 15 $\mu T$ (average value in the Southern Hemisphere), compensating geomagnetic field differences. Corrected PPUTs were linearly interpolated between the North and South directions.}
\label{Fig:PPUT_altitude}
\end{center}
\end{figure}

In the mid-energy range (roughly from 50 GeV to 5 TeV), altitude is no longer such a critical factor in the instrument performance, since the Cherenkov light density is in all sites high enough to produce clear shower images in telescopes within the shower light pool. However, as introduced in Sec. \ref{subsec:altitude},
altitude affects sensitivity in this energy range in several ways: for a given energy, gamma-ray images look more hadron-like at higher altitudes; very close to the shower axis the contribution from particles penetrating to ground level increases with altitude, increasing the level of fluctuations in gamma-ray images and complicating the gamma-hadron separation. In addition, both background suppression and angular resolution are greatly influenced by the telescope multiplicity of the events. At higher altitudes the light pool is smaller, therefore reducing multiplicity for a fixed telescope separation (or reduce the effective area when adjusting telescope separations). The results shown in Fig. \ref{Fig:EffAreas1} and \ref{Fig:PPUT_altitude}, prove that the deteriorated background suppression and lower average telescope multiplicity of measured showers dominate over the higher Cherenkov photon intensity at higher altitudes, showing lower-altitude sites clearly outperform them in this energy range.

In order to evaluate in further details this result, an additional hypothetical site at 500 m altitude (located at Aar, Namibia) was simulated, to test if performance keeps improving with lower altitudes. As shown in Fig. \ref{Fig:DiffSensSouth}, the Aar@500m site shows good performance in the mid-energy range. The benefits of lower-altitude sites described above seem to balance the lower number of measured photons, mainly due to the larger distance of the observatory to the shower maximum. On the other hand, the significantly lower effective area at energies bellow 50 GeV of the (hypothetical) ``Aar@500m" site (shown in Fig. \ref{Fig:EffAreas1}) would deteriorate the Observatory low-energy sensitivity.

As shown in Fig. \ref{Fig:DiffSens40deg}, sites located at slightly higher altitudes, such as Leoncito, are favoured for observations at moderate zenith angles (40 deg), improving sensitivity in the mid-energy range. The shower development at higher zenith angles increases the projected shower light-pool size along with the average shower maximum altitude\footnote{The location of the shower maximum in the atmosphere corresponds to the depth of maximum development, X$_{max}$ (in [g cm$^{-2}$] from the top of the atmosphere)} (H$_\mathrm{max}$). At lower altitude sites, where the distance to H$_\mathrm{max}$ is larger, the loss in Cherenkov photon density surpasses the gain from the improved collection area, decreasing average telescope multiplicity. Taking into account that the average zenith angle for future CTA observations will very likely be around 30 deg, the accepted range in altitudes for such an array of IACTs can be considered rather wide, between $\sim$ 1600 and 2500 m.

As described in Sec. \ref{subsec:gm}, the geomagnetic field bends charged particle trajectories separating positive and negative charges in air showers, leading to a distortion of the Cherenkov light pool on the ground and of the camera image shapes, introducing additional uncertainties in the reconstructed shower parameters. These effects mainly influence the gamma-hadron separation quality (e.g. gamma-ray showers look more hadron-like), the angular (shower image major axis may be slightly shifted and rotated) and energy reconstruction (average Cherenkov light density is not just a function of impact parameter anymore).

\begin{figure}
\begin{center}
\centering\includegraphics[width=0.49\linewidth]{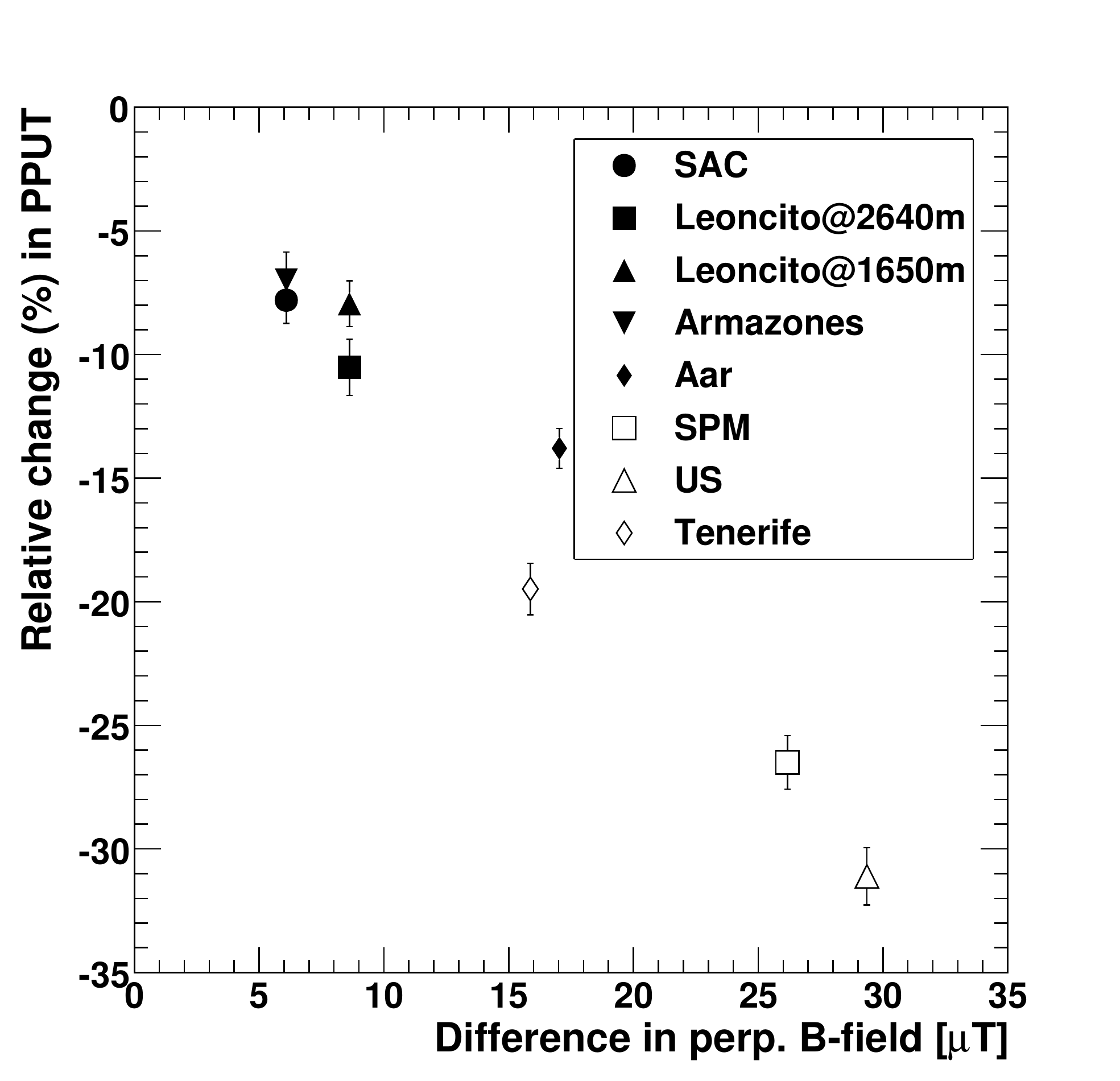}
\centering\includegraphics[width=0.49\linewidth]{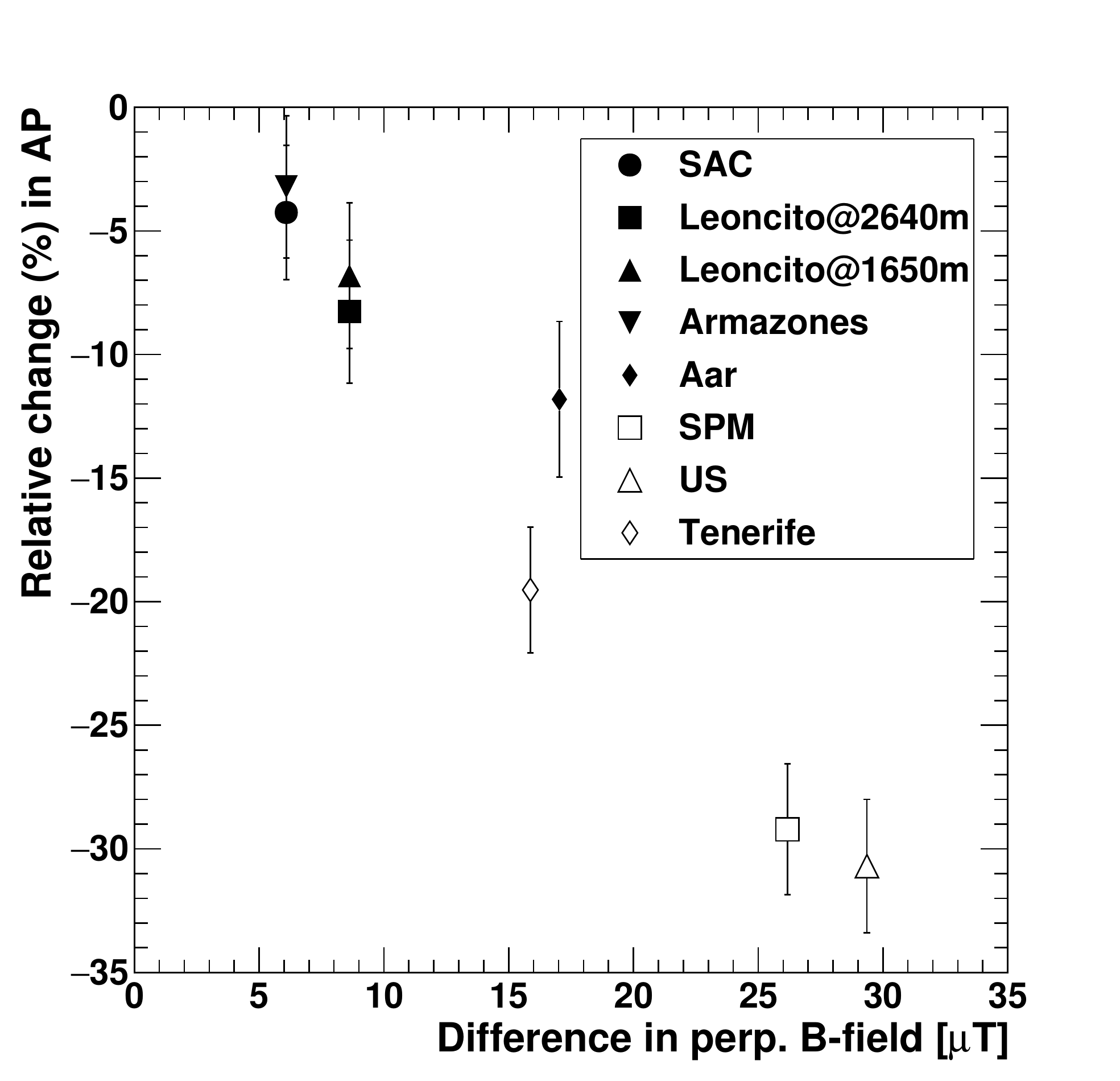}
\caption{Relative difference in PPUT (left) and in AP (right) (see section \ref{sec:sciencePerformance} for definitions) vs perpendicular B-field ($\vec{B}_{\bot}$). The point for a given site is obtained from the comparison of the performance for North and South pointing, which correspond to different values of $\vec{B}_{\bot}$. Northern and Southern site candidates are included in this figure. The relative difference is defined as (PPUT1$-$PPUT2)/(0.5$\times$(PPUT1+PPUT2)), where PPUT1 (PPUT2) corresponds to the pointing direction with the highest (lowest) orthogonal magnetic field (equivalent for the relative difference in AP).}
\label{Fig:PPUT_B}
\end{center}
\end{figure}

Figure \ref{Fig:PPUT_B} and \ref{Fig:PPUT_altitude} (left panel) reveal significant differences in the array performance between different pointing directions. By observing showers with directions close to parallel to the geomagnetic field lines (telescopes pointing to the north for sites in the Southern Hemisphere) performance is significantly better with respect to observations in directions at larger angles to the field lines (up to 10\% higher PPUT for CTA-S). Sensitivity and angular resolution of the sites in South America are inferior to those in Southern Africa, at equal altitude, mainly because the different inclination of the GF results in larger average field intensities perpendicular to the shower direction at small zenith angles in South America. Note the considered zenith angle of 20 deg and the north pointing direction is close to the worst case scenario for the Northern Hemisphere sites, resulting in showers propagating almost perpendicular to the geomagnetic field lines, therefore showing larger differences in PPUT (between 15 and 30\%). All Northern Hemisphere sites studied will be affected, on average, by similar geomagnetic fields. For larger zenith angles differences are smaller.

At the highest energies, above $\sim$5 TeV, the sensitivity is limited by the collection area, which is larger at low-altitude sites (see section \ref{subsec:altitude}).

\subsection{Performance at increased night-sky background levels}
\label{subsec:nsb_results}

\begin{figure}[ht]
\begin{center}
\centering\includegraphics[width=0.6\linewidth]{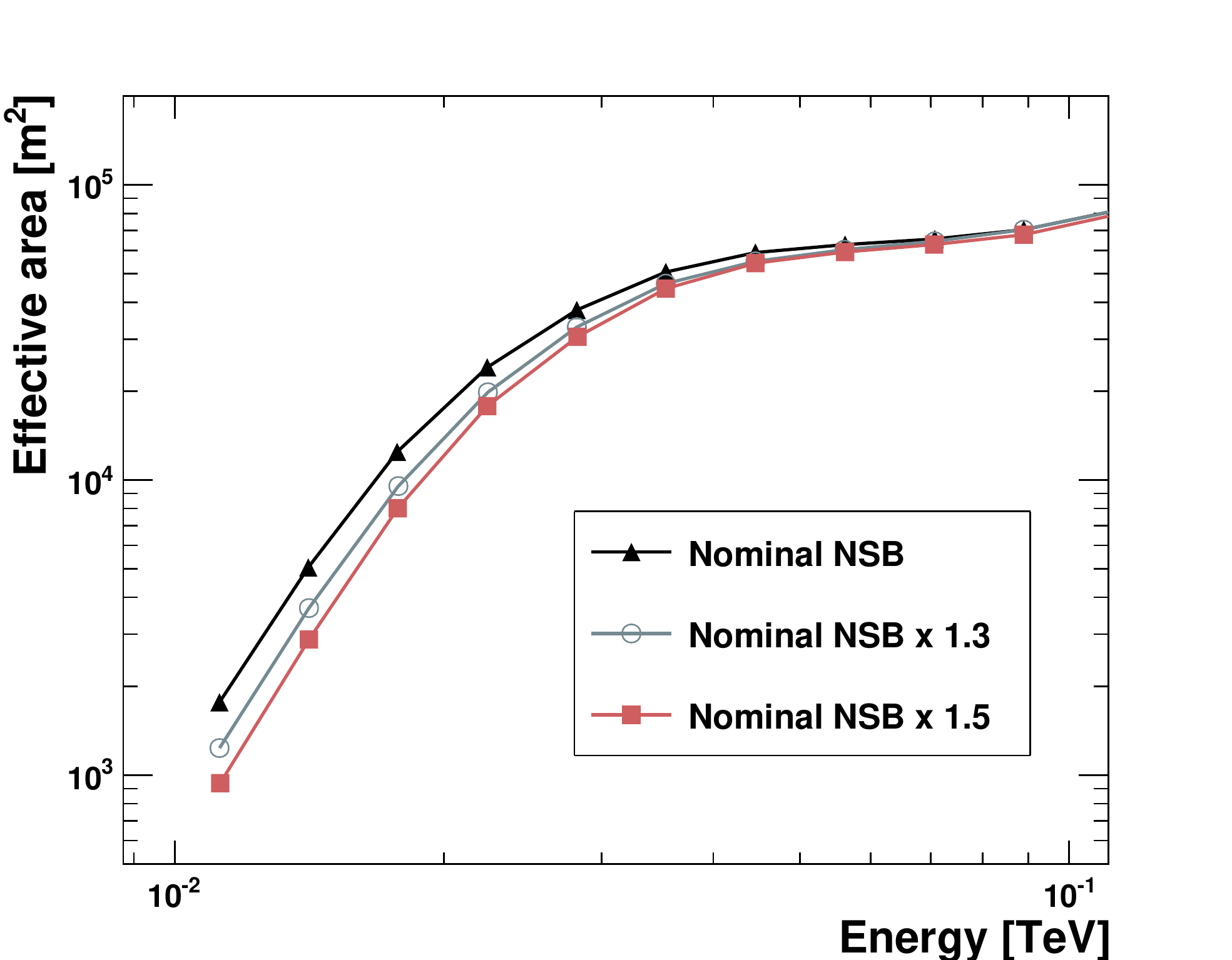}
\caption{Effective areas vs true (simulated) energy in the threshold region for different levels of night-sky background light for the Leoncito site. Very loose cuts have been applied in the analysis for this figure, requiring the successful reconstruction of direction and energy only.
}
\label{Fig:EffAreas2}
\end{center}
\end{figure}

Higher night-sky background levels are expected at some sites due to increased anthropogenic night illumination levels, but also at all sites for observations towards bright regions in the sky (e.g.~within several degrees of the Galactic Plane). As briefly discussed in Sec. \ref{subsec:nsb}, the level of NSB light affects the performance mainly in the threshold region. Higher NSB produces higher accidental rates, requiring increased trigger thresholds. It also lowers the signal-to-noise ratio of measured shower images, leading to a loss of low-energy events and reconstruction quality. 

Figure \ref{Fig:EffAreas2} shows the impact of increasing the nominal NSB level by 30\% and 50\% on the effective area for the Leoncito site. The effect, negligible above 100 GeV, reduces the effective area between 10--20\% at 40 GeV, becoming more significant for lower energies (50--70\% at 25 GeV). These results show that the NSB effect, as compared to the dependence on the site altitude and the geomagnetic field, can be considered of second order, with a marginal effect ($<$5\%) on calculated PPUT values, although quite significant near the energy threshold.

\section{Conclusions}

Although the performance of the CTA candidate sites differs significantly, showing a significant dependence on site altitude and the geomagnetic field intensity, all studied sites hosting a telescope layout such as the one proposed by the CTA Consortium would improve on the sensitivity of current generation of IACTs by a factor 5--10 (energy-dependent).

The study presented here represents the largest-scale simulation ever performed to study the effect of site-dependent parameters on IACTs performance. Results only account for the effect of parameters related to the scientific performance of the Observatory, and do not take into account average annual observation times or logistic arguments which, in the end, may have a greater influence on the final selection.

The best overall performance for a CTA-like observatory is expected for sites at around 1700 m site altitude, with an acceptable range of altitudes between 1600 and 2500 m (see Fig. \ref{Fig:PPUT_altitude}, right panel). Higher altitudes would improve performance below $\sim$ 50 GeV, while significantly decreasing performance above that energy, with an overall differential sensitivity loss of $\sim$ 15\% at 3600 m. Lower altitudes would reduce the low energy performance of the instrument.

The geomagnetic field intensity differences between sites must be also taken into account, as an increase of $\vec{B}_{\bot} \sim$ 10 $\mu$T reduces overall performance a 10\% (both in angular resolution and sensitivity).

Concerning the effect of the NSB, it should be considered of second order, with a marginal effect on the overall array performance (50\% more NSB would decrease less than a 5\% PPUT) and negligible above 100 GeV, although quite significant near the energy threshold.

In the meantime, the CTA site selection process is progressing rapidly: in late March 2015 two sites for each Hemisphere were shortlisted based on the input from the CTA Consortium on annual available observing time, science performance (this study), risks, and cost. During July 2015, detailed contract negotiations started in order to host CTA on the European Southern Observatory (ESO), Paranal site in Chile and at the Instituto de Astrofisica de Canarias (IAC), Roque de los Muchachos Observatory in La Palma, Spain. In September 2016 IAC and the CTA GmbH signed the hosting agreement, while the negotiations with ESO are in an advance state.

\subsubsection*{Acknowledgments}
We gratefully acknowledge support from the agencies and organizations 
listed under Funding Agencies at this website: http://www.cta-observatory.org/ and the European Commission through the contract \#653477.

\bibliography{references.bib}

\begin{thebibliography}{10}
\expandafter\ifx\csname url\endcsname\relax
  \def\url#1{\texttt{#1}}\fi
\expandafter\ifx\csname urlprefix\endcsname\relax\def\urlprefix{URL }\fi
\expandafter\ifx\csname href\endcsname\relax
  \def\href#1#2{#2} \def\path#1{#1}\fi

\bibitem{CTA_concept}
M.~{Actis}, et~al., {Design concepts for the Cherenkov Telescope Array CTA},
  Experimental Astronomy 32 (2011) 193--316.
\newblock \href {http://arxiv.org/abs/1008.3703} {\path{arXiv:1008.3703}},
  \href {http://dx.doi.org/10.1007/s10686-011-9247-0}
  {\path{doi:10.1007/s10686-011-9247-0}}.

\bibitem{CTAICRC2015}
{CTA Consortium}, {Contributions to the 34th International Cosmic Ray
  Conference (ICRC2015)}, ArXiv e-prints:~\href
  {http://arxiv.org/abs/1508.05894} {\path{arXiv:1508.05894}}.

\bibitem{LSTgamma}
D.~{Mazin}, J.~{Cortina}, M.~{Teshima}, t.~{CTA Consortium}, {Large Size
  Telescope Report}, ArXiv e-prints:~\href {http://arxiv.org/abs/1610.04403}
  {\path{arXiv:1610.04403}}.

\bibitem{MSTgamma}
G.~{P{\"u}hlhofer}, f.~t. {CTA Consortium}, {The Medium Size Telescopes of the
  Cherenkov Telescope Array}, ArXiv e-prints:~\href
  {http://arxiv.org/abs/1610.02899} {\path{arXiv:1610.02899}}.

\bibitem{SCTgamma}
W.~{Benbow}, A.~N. {Otte}, {for the SCT and CTA Consortiums}, {Status of the
  Schwarzchild-Couder Medium-Sized Telescope for the Cherenkov Telescope
  Array}, ArXiv e-prints:~\href {http://arxiv.org/abs/1610.03865}
  {\path{arXiv:1610.03865}}.

\bibitem{Montaruli:2015}
T.~{Montaruli}, G.~{Pareschi}, T.~{Greenshaw}, {The small size telescope
  projects for the Cherenkov Telescope Array}, ArXiv e-prints:~\href
  {http://arxiv.org/abs/1508.06472} {\path{arXiv:1508.06472}}.

\bibitem{Hassan-2015}
T.~{Hassan}, et~al., {for the CTA Consortium}, {Second large-scale Monte Carlo
  study for the Cherenkov Telescope Array, in: Proc, 34th ICRC, The Hague,
  2015}, ArXiv e-prints:~\href {http://arxiv.org/abs/1508.06075}
  {\path{arXiv:1508.06075}}.

\bibitem{weather_sims}
S.~Vincent, f.~t. C. T.~A. Consortium, {Atmospheric considerations for the CTA
  site search, in: Proc, AtmoHEAD workshop, Saclay, 2013. }, ArXiv
  e-prints:~\href {http://arxiv.org/abs/1403.5075} {\path{arXiv:1403.5075}}.

\bibitem{APP_CTA_MC}
K.~{Bernl{\"o}hr}, et~al., {Monte Carlo design studies for the Cherenkov
  Telescope Array}, Astroparticle Physics 43 (2013) 171--188.
\newblock \href {http://arxiv.org/abs/1210.3503} {\path{arXiv:1210.3503}}.

\bibitem{MC_ICRC:2013}
K.~{Bernl{\"o}hr}, et~al. {for the CTA Consortium}, {Progress in Monte Carlo
  design and optimization of the Cherenkov Telescope Array, in: Proc, 33th
  ICRC, Rio de Janeiro, 2013}, ArXiv e-prints:~\href
  {http://arxiv.org/abs/1307.2773} {\path{arXiv:1307.2773}}.

\bibitem{MC_ICRC_site:2015}
G.~{Maier}, et~al., {for the CTA Consortium}, {Monte Carlo Performance Studies
  of Candidate Sites for the Cherenkov Telescope Array, in: Proc, 34th ICRC,
  The Hague, 2015}, ArXiv e-prints:~\href {http://arxiv.org/abs/1508.06042}
  {\path{arXiv:1508.06042}}.

\bibitem{impactAtmospherics}
K.~{Bernl{\"o}hr}, {Impact of atmospheric parameters on the atmospheric
  Cherenkov technique}, Astroparticle Physics 12 (2000) 255--268.
\newblock \href {http://arxiv.org/abs/astro-ph/9908093}
  {\path{arXiv:astro-ph/9908093}}, \href
  {http://dx.doi.org/10.1016/S0927-6505(99)00093-6}
  {\path{doi:10.1016/S0927-6505(99)00093-6}}.

\bibitem{concconi:1953}
G.~{Cocconi}, {Influence of the Earth's Magnetic Field on the Extensive Air
  Showers}, Physical Review 95 (1954) 1705--1706.
\newblock \href {http://dx.doi.org/10.1103/PhysRev.95.1705.4}
  {\path{doi:10.1103/PhysRev.95.1705.4}}.

\bibitem{Reyes:2009}
R.~d.~l. Reyes, {Search for Gamma-ray emission from pulsars with the MAGIC
  telescope: Sensitivity studies, data check and data analysis}, {PhD}
  dissertation, Universidad Complutense de Madrid, Atomic and Nuclear Physics
  Department (2009).

\bibitem{corsika}
D.~{Heck}, J.~{Knapp}, J.~N. {Capdevielle}, G.~{Schatz}, T.~{Thouw}, {CORSIKA:
  a Monte Carlo code to simulate extensive air showers}, Forschungszentrum
  Karlsruhe GmbH, {FZKA-6019 (1998)}.

\bibitem{Szanecki_2013}
M.~{Szanecki}, et~al., {Influence of the geomagnetic field on the IACT
  detection technique for possible sites of CTA observatories}, Astroparticle
  Physics 45 (2013) 1--12.
\newblock \href {http://arxiv.org/abs/1302.6387} {\path{arXiv:1302.6387}},
  \href {http://dx.doi.org/10.1016/j.astropartphys.2013.02.002}
  {\path{doi:10.1016/j.astropartphys.2013.02.002}}.

\bibitem{ICRC_2008_GF}
S.~C. {Commichau}, et~al., {Geomagnetic Field Effects on the Imaging Air Shower
  Cherenkov Technique}, International Cosmic Ray Conference 3 (2008)
  1357--1360.
\newblock \href {http://arxiv.org/abs/0709.1251} {\path{arXiv:0709.1251}}.

\bibitem{sitePaper3}
C.~{Fruck}, M.~{Gaug}, J.-P. {Ernenwein}, D.~{Mand{\'a}t}, T.~{Schweizer},
  D.~{H{\"a}fner}, T.~{Bulik}, M.~{Cieslar}, H.~{Costantini}, M.~{Dominik},
  J.~{Ebr}, M.~{Garczarczyk}, E.~{Lorentz}, G.~{Pareschi}, M.~{Pech},
  I.~{Puerto-Gim{\'e}nez}, M.~{Teshima}, {Instrumentation for comparing night
  sky quality and atmospheric conditions of CTA site candidates}, Journal of
  Instrumentation 10 (2015) P04012.
\newblock \href {http://arxiv.org/abs/1501.02156} {\path{arXiv:1501.02156}},
  \href {http://dx.doi.org/10.1088/1748-0221/10/04/P04012}
  {\path{doi:10.1088/1748-0221/10/04/P04012}}.

\bibitem{NSBmarkus}
M.~{Gaug}, {for the CTA Consortium}, {Night Sky Background Analysis for the
  Cherenkov Telescope Array using the Atmoscope instrument, in: Proc, 33th
  ICRC, Rio de Janeiro, 2013}, ArXiv e-prints:~\href
  {http://arxiv.org/abs/1307.3053} {\path{arXiv:1307.3053}}.

\bibitem{Konrad:2008}
K.~{Bernl{\"o}hr}, {Simulation of imaging atmospheric Cherenkov telescopes with
  CORSIKA and sim\_telarray}, Astroparticle Physics 30 (2008) 149--158.
\newblock \href {http://arxiv.org/abs/0808.2253} {\path{arXiv:0808.2253}},
  \href {http://dx.doi.org/10.1016/j.astropartphys.2008.07.009}
  {\path{doi:10.1016/j.astropartphys.2008.07.009}}.

\bibitem{atmModel}
J.~M. {Picone}, et~al., {NRLMSISE-00 empirical model of the atmosphere:
  Statistical comparisons and scientific issues}, Journal of Geophysical
  Research (Space Physics) 107 (2002) 1468.
\newblock \href {http://dx.doi.org/10.1029/2002JA009430}
  {\path{doi:10.1029/2002JA009430}}.

\bibitem{sitePaper1}
R.~D. {Piacentini}, B.~{Garc{\'{\i}}a}, M.~I. {Micheletti}, G.~{Salum},
  M.~{Freire}, J.~{Maya}, A.~{Mancilla}, E.~{Crin{\'o}}, D.~{Mandat},
  M.~{Pech}, T.~{Bulik}, {Selection of astrophysical/astronomical/solar sites
  at the Argentina East Andes range taking into account atmospheric
  components}, Advances in Space Research 57 (2016) 2559--2574.
\newblock \href {http://dx.doi.org/10.1016/j.asr.2016.03.027}
  {\path{doi:10.1016/j.asr.2016.03.027}}.

\bibitem{sitePaper2}
G.~{Tovmassian}, M.-S. {Hernandez}, J.~L. {Ochoa}, J.-P. {Ernenwein},
  D.~{Mandat}, M.~{Pech}, I.~{Plauchu Frayn}, E.~{Colorado}, J.~M. {Murillo},
  U.~{Cese{\~n}a}, B.~{Garcia}, W.~H. {Lee}, T.~{Bulik}, M.~{Garczarczyk},
  C.~{Fruck}, H.~{Costantini}, M.~{Cieslar}, T.~{Aune}, S.~{Vincent},
  J.~{Carr}, N.~{Serre}, P.~{Janecek}, D.~{Haefner}, {Astroclimatic
  Characterization of Vallecitos: A Candidate Site for the Cherenkov Telescope
  Array at San Pedro M{\'a}rtir}, \pasp 128~(3) (2016) 035004.
\newblock \href {http://arxiv.org/abs/1601.02383} {\path{arXiv:1601.02383}},
  \href {http://dx.doi.org/10.1088/1538-3873/128/961/035004}
  {\path{doi:10.1088/1538-3873/128/961/035004}}.

\bibitem{Prod2_SCMST}
T.~{Hassan}, et~al., {for the CTA Consortium}, {Layout design studies for
  medium-sized telescopes within the Cherenkov Telescope Array, in: Proc, 34th
  ICRC, The Hague, 2015}, ArXiv e-prints:~\href
  {http://arxiv.org/abs/1508.06076} {\path{arXiv:1508.06076}}.

\bibitem{dirac-general}
A.~{Tsaregorodtsev}, {Dirac Project}, {DIRAC Distributed Computing Services},
  Journal of Physics Conference Series 513~(3) (2014) 032096.
\newblock \href {http://dx.doi.org/10.1088/1742-6596/513/3/032096}
  {\path{doi:10.1088/1742-6596/513/3/032096}}.

\bibitem{Arrabito-2015}
L.~{Arrabito}, et~al., Prototype of a production system for {Cherenkov
  Telescope Array} with {DIRAC}, Journal of Physics: Conference Series 664~(3)
  (2015) 032001.

\bibitem{evnDisplay}
G.~{Maier}, A short description of an \textit{evnDisplay}-based {CTA} analysis,
  \url{https://znwiki3.ifh.de/CTA/Eventdisplay\%20Software}, accessed:
  2016-09-13.

\bibitem{MARS}
A.~{Moralejo}, et~al., {MARS, the MAGIC Analysis and Reconstruction Software,
  in: Proc, 31st ICRC, {\L}odz, 2009}, ArXiv e-prints:~\href
  {http://arxiv.org/abs/0907.0943} {\path{arXiv:0907.0943}}.

\bibitem{Wood:2014}
M.~{Wood}, et~al., {Monte Carlo studies of medium-size telescope designs for
  the Cherenkov Telescope Array}, Astroparticle Physics 72 (2016) 11--31.
\newblock \href {http://arxiv.org/abs/1506.07476} {\path{arXiv:1506.07476}},
  \href {http://dx.doi.org/10.1016/j.astropartphys.2015.04.008}
  {\path{doi:10.1016/j.astropartphys.2015.04.008}}.

\bibitem{Daum:1997}
A.~{Daum}, {et al.}, {First results on the performance of the HEGRA IACT
  array}, Astroparticle Physics 8 (1997) 1--11.
\newblock \href {http://dx.doi.org/10.1016/S0927-6505(97)00031-5}
  {\path{doi:10.1016/S0927-6505(97)00031-5}}.

\bibitem{hillas}
A.~M. {Hillas}, {Cerenkov light images of EAS produced by primary gamma rays
  and by nuclei}, International Cosmic Ray Conference 3 (1985) 445--448.

\bibitem{Breiman2001}
L.~Breiman, Random forests, Machine Learning 45~(1) (2001) 5--32.

\bibitem{hadronness}
J.~{Albert}, et~al., {Implementation of the Random Forest method for the
  Imaging Atmospheric Cherenkov Telescope MAGIC}, Nuclear Instruments and
  Methods in Physics Research A 588 (2008) 424--432.
\newblock \href {http://arxiv.org/abs/0709.3719} {\path{arXiv:0709.3719}},
  \href {http://dx.doi.org/10.1016/j.nima.2007.11.068}
  {\path{doi:10.1016/j.nima.2007.11.068}}.

\bibitem{LiMa}
T.~P. {Li}, Y.~Q. {Ma}, {Analysis methods for results in gamma-ray astronomy},
  The Astrophysical Journal 272 (1983) 317--324.
\newblock \href {http://dx.doi.org/10.1086/161295} {\path{doi:10.1086/161295}}.

\bibitem{MAGIC_disp}
J.~{Aleksi{\'c}}, et~al., {Search for an extended VHE {$\gamma$}-ray emission
  from Mrk 421 and Mrk 501 with the MAGIC Telescope}, \aap 524 (2010) A77.
\newblock \href {http://arxiv.org/abs/1004.1093} {\path{arXiv:1004.1093}},
  \href {http://dx.doi.org/10.1051/0004-6361/201014747}
  {\path{doi:10.1051/0004-6361/201014747}}.

\bibitem{HESS_disp}
C.-C. {Lu}, {for the H.~E.~S.~S.~Collaboration}, {Improving the H.E.S.S.
  angular resolution using the Disp method}, ArXiv e-prints:~\href
  {http://arxiv.org/abs/1310.1200} {\path{arXiv:1310.1200}}.

\bibitem{VERITAS_disp}
G.~D. {Senturk}, {The Disp Method for Analysing Large Zenith Angle Gamma-Ray
  Data}, International Cosmic Ray Conference 9 (2011) 127.
\newblock \href {http://arxiv.org/abs/1109.6044} {\path{arXiv:1109.6044}},
  \href {http://dx.doi.org/10.7529/ICRC2011/V09/0925}
  {\path{doi:10.7529/ICRC2011/V09/0925}}.

\bibitem{Maxim_cleaning_magic}
M.~{Shayduk}, T.~{Hengstebeck}, O.~{Kalekin}, N.~A. {Pavel}, T.~{Schweizer}, {A
  New Image Cleaning Method for the MAGIC Telescope}, International Cosmic Ray
  Conference 5 (2005) 223.

\bibitem{Maxim_cleaning_cta}
M.~{Shayduk}, {for the CTA Consortium}, {Optimized next-neighbor image cleaning
  method for Cherenkov Telescopes}, ArXiv e-prints:~\href
  {http://arxiv.org/abs/1307.4939} {\path{arXiv:1307.4939}}.

\bibitem{modelAnalysis}
M.~{Holler}, et~al., {for the H.~E.~S.~S.~collaboration}, {Photon
  Reconstruction for H.E.S.S. Using a Semi-Analytical Shower Model, in: Proc,
  34th ICRC, The Hague, 2015}, ArXiv e-prints:~\href
  {http://arxiv.org/abs/1509.02896} {\path{arXiv:1509.02896}}.

\end{thebibliography}

\doclicenseThis

\end{document}